\documentclass[twocolumn]{aastex62}

\usepackage{tcolorbox}
\usepackage{amsmath}

\newcommand{\summary}[1]{}

\newcommand{\changed}[1]{#1}

\submitjournal{ApJ}

\received{June 24, 2019}
\revised{August 2, 2019}
\accepted{August 3, 2019}

\begin{document}

\title{PHASE-APODIZED-PUPIL LYOT CORONAGRAPHS FOR ARBITRARY TELESCOPE PUPILS}

\correspondingauthor{Emiel H. Por}
\email{por@strw.leidenuniv.nl}

\author{Emiel H. Por}
\affiliation{Leiden Observatory, Leiden University, P.O. Box 9513, 2300 RA Leiden, The Netherlands}

\newcommand{\vect}[1]{\mathbf{#1}}
\newcommand{\fourier}[1]{\mathcal{F}\{#1\}}
\newcommand{\invfourier}[1]{\mathcal{F}^{-1}\{#1\}}
\newcommand{\prop}[2][\lambda]{\mathcal{P}_{#1}\{#2\}}
\newcommand{\invprop}[2][\lambda]{\mathcal{P}_{#1}^{-1}\{#2\}}
\newcommand{\Real}[1]{\mathfrak{R}\left\{ {#1} \right\}}
\newcommand{\Imag}[1]{\mathfrak{I}\left\{ {#1} \right\}}

\begin{abstract} 
The phase-apodized-pupil Lyot coronagraph (PAPLC) is a pairing of the apodized-pupil Lyot coronagraph (APLC) and the apodizing phase plate (APP) coronagraph. We describe a numerical optimization method to obtain globally-optimal solutions for the phase apodizers for arbitrary telescope pupils, based on the linear map between complex-amplitude transmission of the apodizer and the electric field in the post-coronagraphic focal plane. PAPLCs with annular focal-plane masks and point-symmetric dark zones perform analogous to their corresponding APLCs. However with a knife-edge focal-plane mask and one-sided dark zones, the PAPLC yields inner working angles as close as $1.4\lambda/D$ at contrasts of $10^{-10}$ and maximum post-coronagraphic throughput of $>75\%$ for telescope apertures with central obscurations of up to $30\%$. We present knife-edge PAPLC designs optimized for the VLT/SPHERE instrument and the LUVOIR-A aperture. These designs show that the knife-edge PAPLC retains its performance, even for realistic telescope pupils with struts, segments and non-circular outer edges.
\end{abstract}


\section{Introduction}
\label{sec:introduction}

In the last few decades, we have seen tremendous advances in the field of exoplanets. Initiated by the discovery of the first planet orbiting another main-sequence star by \cite{mayor1995jupiter}, we now know that most stars harbor a companion in the habitable zone \citep{borucki2011characteristics}. The majority of planets are detected using indirect methods, such as radial velocity \citep{mayor1995jupiter} and transits \citep{Charbonneau2000, Henry2000}. For the brightest stars with transiting planets, spectral characterisation is possible during the transit itself. Longer period planetary transits require precise ephemerides and are limited by the decreasing frequency of observed transits. Direct imaging of these planetary systems provides a way for the detection and characterization of the atmospheres, including variability induced by the rotational modulation of cloud and weather systems and the discovery of liquid water surfaces through glints off liquid surface detectable with polarization. 

With the advent of extreme adaptive optics systems, such as VLT/SPHERE \citep{beuzit2008sphere}, Gemini/GPI \citep{Macintosh2008gpi}, Clay/MagAO-X \citep{Close2012SPIEMagAO, Males2014MagAO}, and Subaru/SCExAO \citep{Jovanovic2015scexao}, and dedicated space-based instrumentation, such as WFIRST/CGI \citep{spergel2013wfirst} and HabEx \citep{mennesson2016habex}, spatially-resolved imaging of exoplanets has started to become a reality. An optical system known as a coronagraph filters out the light from the on-axis star, while letting through the light from off-axis sources, such as that from faint companions or debris disks. This permits analysis of the off-axis light directly, without being overwhelmed by the on-axis star, and therefore easier chemical characterization of the material orbiting the star. Coronagraphs are \changed{both currently used and planned} for both future and current space- and ground-based systems.

Many families of coronagraphs have been developed over the years. Among the simplest are the pupil-plane coronagraphs. These coronagraphs apodize the light only in a single pupil plane. The pattern of apodization is designed in such a way as to generate a dark region in the focal plane. Note that, as both on- and off-axis light is apodized in the same way, the apodization pattern must be as minor as possible as to not block too much of the light from the companion or disk. Generally during the design process of such a coronagraph, the throughput is maximized while simultaneously constraining the stellar intensity in the dark zone. Pupil-plane coronagraphs can be separated into two types:
\begin{itemize}
    \item \emph{Shaped pupil coronagraphs (SPC).} These coronagraphs apodize the pupil with \changed{a binary} amplitude pattern. \changed{Amplitude apodization initially started off as grey-scale \citep{slepian1965analytic}, but has since changed to} binary \citep{kasdin2003extrasolar}, \changed{as} \cite{carlotti2011optimal} \changed{showed} that convex optimization of \changed{a gray-scale} apodizer yields a globally-optimal binary amplitude mask. SPCs can only create dark zones with point symmetry: \changed{as the Fourier transform of a real function is Hermitian, any amplitude-apodized pupil, either binary or gray-scale, inherently has a point-symmetric point spread function (PSF)}.
    \item \emph{Apodizing phase plate coronagraphs (APP).} These coronagraphs apodize the pupil with a phase-only mask \citep{codona2006high,snik2012vector,otten2017sky}. Early designs used Fourier iteration techniques \citep{codona2006high} to find a valid phase pattern. Currently globally-optimal phase patterns can be found using direct convex optimization \citep{por2017optimal}. APPs can create dark zones with \changed{or} without point symmetry.
\end{itemize}

While it may seem that combining both phase and amplitude apodizing in a pupil-plane coronagraph might yield coronagraphs with higher throughput than either SPCs and APPs, this is not the case. \cite{por2017optimal} shows that global optimization of a complex-amplitude pupil-plane apodizer will always yield a phase-only apodizer. A corollary is that an APP coronagraph will always outperform a SPC, barring implementation details, as the solution space for SPCs is a subset of the solution space for pupil-plane coronagraphs with a complex-amplitude apodizer. That is, for a fixed telescope pupil shape, dark zone geometry and contrast requirement, the optimal APP will have the same or a higher throughput compared to the optimal SPC. In practice however, for point-symmetric dark zones the gain in throughput is usually minimal, except when the design requirements are so demanding that the throughput is already low for both the SPC and APP coronagraphs \citep{por2017optimal}.

The sheer simplicity of the optical layout of pupil-plane coronagraphs has led to their widespread use in high-contrast imaging instruments \citep{otten2017sky, doelman2017patterned, currie2018laboratory}. However this simple optical layout also implies worse performance compared to coronagraphs with a more complicated optical layout, due to their more limited design freedom. Because of this, the SPC is often combined with a Lyot stage downstream of the apodizer \citep{soummer2004apodized, zimmerman2016shaped}. \changed{A Lyot stage consists of a focal-plane mask, which apodizes part of the point spread function, and a pupil-plane mask, called a Lyot-stop mask, that further filters out the residual stellar light. An SPC combined with a Lyot stage is called an} Apodized Pupil Lyot Coronagraph (APLC). The added Lyot stage has the effect of reducing the inner working angle and allowing deeper design contrasts. The APLC is able to achieve space-based contrasts at reasonable inner working angles and throughput, making it a baseline coronagraph to which other, more complicated coronagraph designs are compared \citep{pueyo2017luvoir,riggs2017shaped}.

The success of the APLC leads us to the question: what is the performance of a phase-apodized-pupil Lyot coronagraph (PAPLC)? In Section~\ref{sec:optimization} we will outline the numerical optimization method for designing a PAPLC. We will distinguish two types of PAPLCs: one with an annular focal-plane mask and point-symmetric dark zones, and one with a knife-edge focal-plane mask and one-sided dark zones. We will perform a study for the parameter space for simplified telescope pupils for each type in Section~\ref{sec:parameter_study:point-symmetric}~and~\ref{sec:parameter_study:one-sided} respectively. To demonstrate the PAPLC for realistic telescope pupils we show designs for the VLT/SPHERE instrument and LUVOIR-A telescope in Section~\ref{sec:case_studies}. Finally, we will conclude with Section~\ref{sec:conclusions}.

\section{Overview of the numerical optimization problem}
\label{sec:optimization}

In this section we will outline the optimization procedure for PAPLCs. This procedure is based on convex optimization and modifies that of \cite{por2017optimal}, where convex optimization is used for optimizing APPs. We will start by formally defining the optimization problem. Then we will convexify this problem to make global optimization more efficient. Furthermore, we will study how symmetries can be included in the optimization and how these affect the optimal phase pattern. Finally, we discuss how to constrain the tip-tilt of the apodizer in a way that keeps the optimization problem convex.

\subsection{Problem definition}
\label{sec:optimization:problem}

\begin{figure}
    \plotone{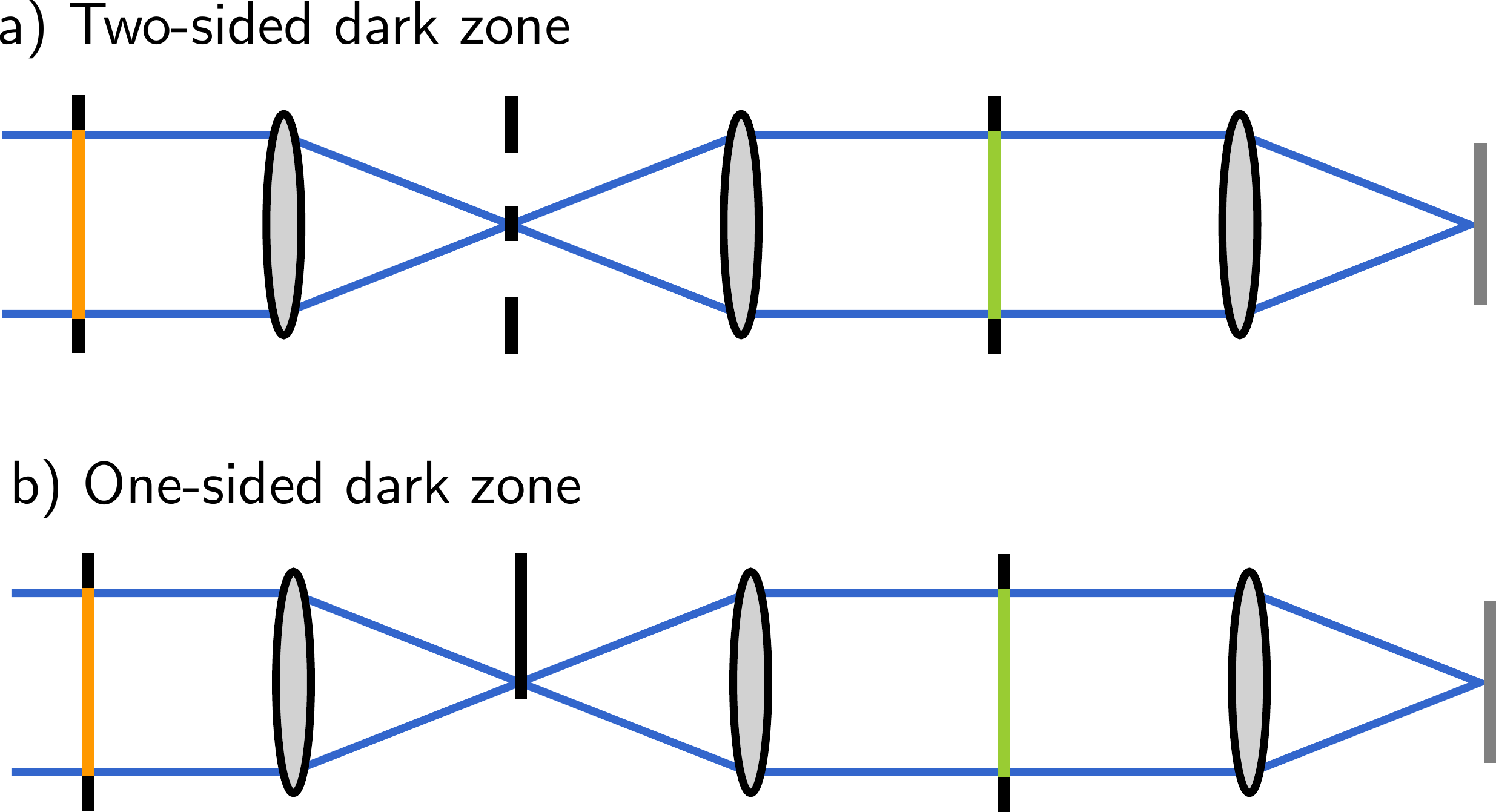}
    \caption{The optical layout of the PAPLC with \emph{a)} point-symmetric dark zones, and \emph{b)} one-sided dark zones follows a standard Lyot-style optical setup. The focal-plane mask for point-symmetric dark zones is annular, while it is a knife edge for the one-sided dark zone. In this study we optimize the pre-apodizer (in orange), viewing the parameters of the focal-plane mask and the Lyot stop (in green) as hyperparameters.}
    \label{fig:optical_layout}
\end{figure}

The optical layout of the PAPLC is shown schematically in Figure~\ref{fig:optical_layout}. While joint optimization of the focal-plane mask and Lyot stop is in principle possible, we will restrict ourselves in this study to parameterized focal-plane masks and Lyot stops only. Their parameters will be viewed as hyperparameters on the optimization problem for finding the optimal apodizer. In this study, the number of hyperparameters is limited, and brute-force optimization is used to optimize them at an acceptable performance cost. More advanced black-box global optimizers, such as Bayesian optimization approaches \citep{kushner1964new,snoek2012practical} or Monte-Carlo techniques \citep{fogarty2018optimal}, can be used if more hyperparameters are required.

\begin{figure}
    \plotone{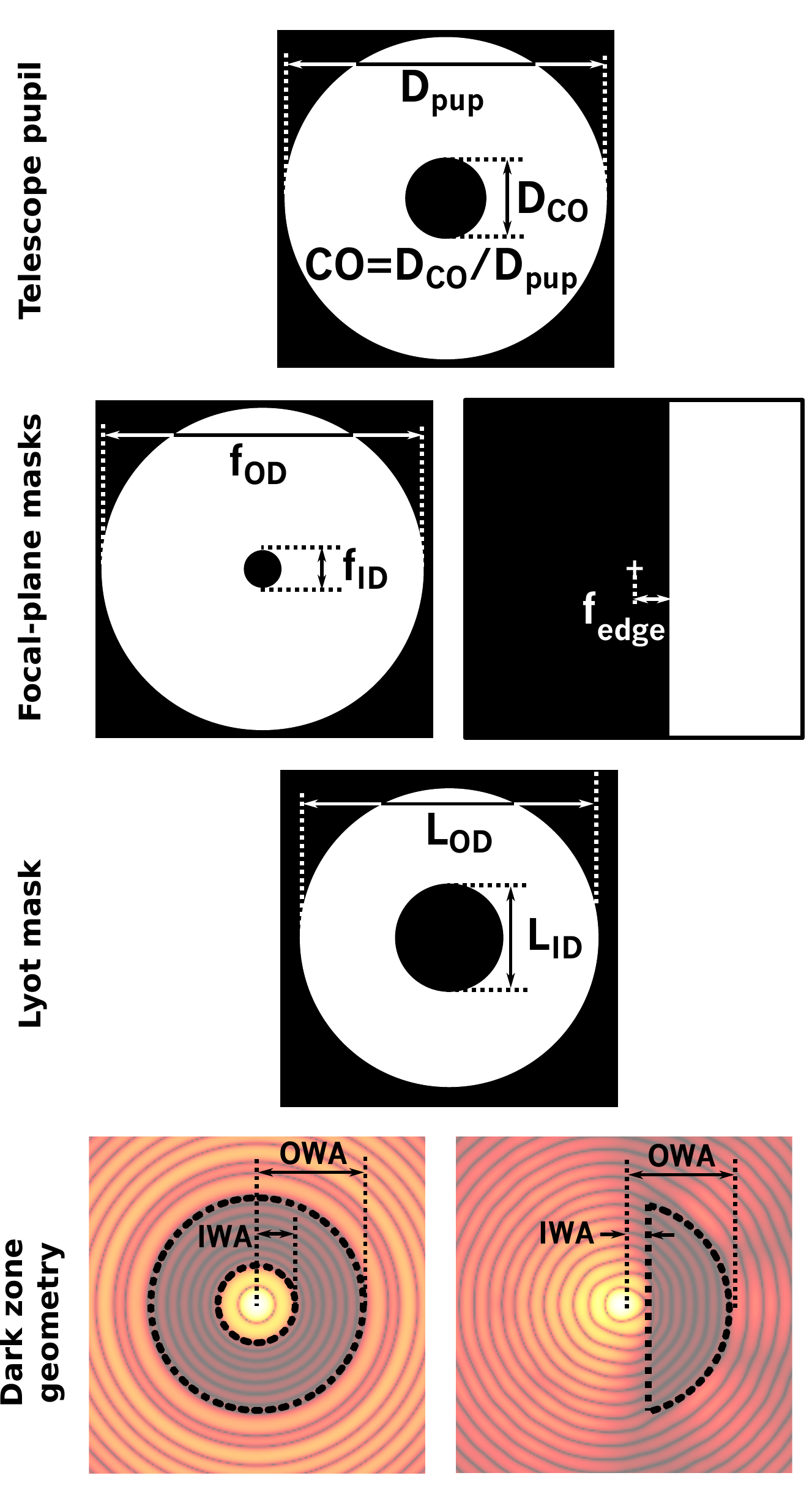}
    \caption{The definition of all masks used in this work. These masks are used for the parameter study in Sections~\ref{sec:parameter_study:point-symmetric}~and~\ref{sec:parameter_study:one-sided}. Centered masks are used for both point-symmetric and one-sided dark zones. The left-justified masks are for two-sided dark zones, while the right-justified masks are used for one-sided dark zones. In general though, arbitrary telescope pupils, Lyot masks, focal-plane masks and dark zone geometries can be used with a PAPLC.}
    \label{fig:mask_definitions}
\end{figure}

Additionally, while many types of focal-plane mask designs are possible, we restrict ourselves in this study to either annular focal-plane masks for point-symmetric dark zones, or an offset knife-edge focal-plane mask for one-sided dark zones. For our parameter studies in Sections~\ref{sec:parameter_study:point-symmetric}~and~\ref{sec:parameter_study:one-sided} we will use simplified apertures. There we will use a circularly-obscured telescope pupil and an annular Lyot stop. Furthermore, we will solely use annular dark zones for the point-symmetric dark zones, and D-shaped dark zones as one-sided dark zones. All parameters for the telescope pupil, focal-plane mask, Lyot stop and dark zone geometry are shown schematically in Figure~\ref{fig:mask_definitions}.

We will use aperture photometry as the main metric for coronagraph performance, and follow \cite{ruane2018review} for our definitions. Here we give a short summary of these definitions for completeness.

We define $\eta_0$ as the encircled energy within a circle with a radius of $0.7\lambda/D$ of a normalized PSF generated by the optical system without any coronagraphic masks, so with no \changed{apodizer mask,} focal-plane mask or Lyot stop mask. This PSF is normalized such that the total power equals one. We define $\eta_p(\vect{k}, \lambda)$ as the encircled energy within a circle with a radius of $0.7\lambda/D$ centered around $\vect{k}$, of the planetary, off-axis PSF, where the planet is located at $\vect{k}$, through the coronagraphic optical system. We define $\eta_s(\vect{k}, \lambda)$ as the encircled energy within a circle with a radius of $0.7\lambda/D$ centered around $\vect{k}$, of the stellar, on-axis image through the coronagraphic optical system. We can now define the throughput $T(\vect{k}, \lambda)$ as the ratio between encircled energies of the non-coronagraphic PSF and the off-axis coronagraphic PSF:
\begin{equation}
    T(\vect{k}, \lambda) = \eta_p(\vect{k}, \lambda) / \eta_0.
\end{equation}
The raw contrast $C(\vect{k}, \lambda)$ is defined as the ratio between stellar and planetary encircled energies:
\begin{equation}
    C(\vect{k}, \lambda) = \eta_s(\vect{k}, \lambda) / \eta_p(\vect{k}, \lambda).
\end{equation}
The design raw contrast $C_\mathrm{design}$ is defined as the maximum raw contrast in the dark zone $D$ over the whole spectral band:
\begin{equation}
    C_\mathrm{design} = \max_{\vect{k}\in D, \lambda\in[\lambda_-, \lambda_+]} C(\vect{k}),
\end{equation}
where $\lambda_-, \lambda_+$ are the minimum and maximum wavelength in the spectral band. Finally we define the inner working angle $\mathit{IWA}$ as the smallest angular separation for which the throughput is larger than half of its maximum value for the whole spectral band:
\begin{equation}
    \mathit{IWA} = \min_{\{\vect{k} : T(\vect{k},\lambda) > \frac12 \max_{\vect{k}}. T(\vect{k},\lambda)\}, \lambda\in[\lambda_-, \lambda_+]}|\vect{k}|
\end{equation}

We can now define the optimization problem for the PAPLC. We try to maximize the throughput of the planet while simultaneously constraining the raw contrast in the dark zone. The phase pattern $\phi(\vect{x})$ can vary across the aperture. As the throughput $T(\vect{k}, \lambda)$ varies across the field of view and as function of wavelength across the spectral band, we take the maximum attained throughput at the center wavelength $\lambda_0$ as a measure for the overall throughput. The optimization problem is given by:
\begin{subequations}
\begin{align}
\underset{\phi(\vect{x})}{\operatorname{maximize}} &~~~ \max_\vect{k} T(\vect{k}, \lambda_0) \label{eq:full_optim_obj} \\
\operatorname{subject~to} &~~~ \eta_s(\vect{k}, \lambda) < \eta_p(\vect{k}, \lambda) \cdot 10^{-c(\vect{k})}\\
\nonumber&~~~~~~~~~~~~~~~~~~~~~\forall~\vect{k} \in D~\forall~\lambda \in [\lambda_-, \lambda_+], \label{eq:full_optim_constr}
\end{align}
\end{subequations}
where $10^{-c(\vect{k})}$ is the design contrast in the dark zone, $\vect{x}$ is a position in the pre-apodizer, $\vect{k}$ is a position in the post-coronagraphic focal plane, $D$ is the dark zone, $\lambda$ is the wavelength of the light, and $[\lambda_-, \lambda_+]$ is the spectral bandwidth for which we want to optimize.

\subsection{Simplification and convexification}
\label{sec:optimization:convexification}

This optimization problem is non-convex. This means that there could be many local optima and ensuring that the found solution is globally optimal requires a full search of the parameter space. We often prefer convex optimization problems, as they only permit only a single local optimum (which is then also globally optimal). This makes solving convex optimization problems much easier than non-convex problems. \changed{In order to convexify our non-convex optimization problem, we need to simplify it quite a bit.}

We will discard the aperture photometry methodology in the optimization procedure. \changed{This will help us to convexify the objective function later on and will simplify the notation.} We will still evaluate all designs using aperture photometry. This yields for the optimization problem:
\begin{subequations}
\begin{align}
\underset{\phi(\vect{x})}{\operatorname{maximize}} &~~~ |E_{\mathrm{noncoro}, \lambda_0}(\vect{0})|^2\\
\operatorname{subject~to} &~~~ |E_{\mathrm{coro},\lambda}(\vect{k})|^2 < 10^{-c(\vect{k})} |E_{\mathrm{noncoro}, \lambda}(\vect{k})|^2\\
\nonumber&~~~~~~~~~~~~~~~~~~~~~\forall~\vect{k} \in D~\forall~\lambda \in [\lambda_-, \lambda_+],
\end{align}
\end{subequations}
where $E_{\mathrm{coro},\lambda}(\vect{k})$ is the on-axis PSF at wavelength $\lambda$ and $E_{\mathrm{noncoro},\lambda}(\vect{k})$ is the on-axis PSF without the focal-plane mask but with the \changed{apodizer and} Lyot stop mask in the optical system:
\begin{subequations}
\begin{align}
    E_{\mathrm{coro},\lambda}(\vect{k}) &= \prop{L(\vect{x}) \invprop{M(\vect{k}) \prop{E_\mathrm{pup}(\vect{x})}}}, \\
    E_{\mathrm{noncoro},\lambda}(\vect{k}) &= \prop{L(\vect{x}) E_\mathrm{pup}(\vect{x})}, \\
    E_{\mathrm{pup}}(\vect{x}) &= A(\vect{x}) \exp{i\phi(\vect{x})},
\end{align}
\end{subequations}
where $A(\vect{x})$ is the telescope pupil, $M(\vect{k})$ is the focal-plane mask, $L(\vect{x})$ is the Lyot stop, $\prop{\cdot}$ is the propagation operator that propagates an electric field from a pupil plane to a focal plane given a wavelength of $\lambda$ and $\invprop{\cdot}$ is the inverse of this operator, propagating an electric field from a focal plane to a pupil plane.

This simplification makes the optimization more tractable, but not yet convex. We change the complex phase exponential $\exp{i\phi(\vect{x})}$ into the complex amplitude $X(\vect{x})+iY(\vect{x})$, so that
\begin{equation}
    E_\mathrm{pup}(\vect{x}) = A(\vect{x}) (X(\vect{x}) + iY(\vect{y})),
\end{equation}
and add the phase-only constraint
\begin{equation}
    X^2(\vect{x}) + Y^2(\vect{x}) = 1
\end{equation}
to the optimization problem. This additional constraint requires the amplitude of the now complex-amplitude apodizer transmission to be one.

Furthermore we can remove the piston symmetry from the optimization problem: the \changed{problem} is invariant under the transformation $S: \phi(\vect{x}) \to \phi(\vect{x}) + \alpha$, where $\alpha$ is any arbitrary constant. \changed{So when we have found a solution $\hat{\phi}(\vect{x})$, we know that $S \hat{\phi}(\vect{x}) = \hat{\phi}(\vect{x}) + \alpha$ is also a solution of the problem. This means that the solution to the problem is non-unique and the problem therefore non-convex.} We remove this symmetry by maximizing the real part of the non-coronagraphic electric field, rather than its absolute value. \changed{The} choice of maximizing the real part, instead of any other linear combination of real and imaginary part is arbitrary. \changed{The removal of this symmetry alone does not guarantee a unique solution in general; it only removes a source of non-convexity from the problem.} The optimization problem now reads:
\begin{subequations}
\begin{align}
\underset{X(\vect{x}), Y(\vect{x})}{\operatorname{maximize}} &~~~ \Real{E_{\mathrm{noncoro}, \lambda_0}(\vect{0})}\\
\operatorname{subject~to} &~~~ |E_{\mathrm{coro},\lambda}(\vect{k})|^2 < 10^{-c(\vect{k})} |E_{\mathrm{noncoro}, \lambda}(\vect{k})|^2\\
\nonumber&~~~~~~~~~~~~~~~~~~~~~\forall~\vect{k} \in D~\forall~\lambda \in [\lambda_-, \lambda_+] \\
&~~~ X^2(\vect{x}) + Y^2(\vect{x}) = 1\forall \vect{x}.
\end{align}
\end{subequations}

At this point the objective function is fully linear and therefore convex, and the first constraint is quadratic but convex as well. The only remaining source of non-convexity stems from the phase-only constraint on the complex-amplitude apodizer transmission. Similar to \cite{por2017optimal} we allow the apodizer to vary not only in phase, but also in amplitude. This convexifies the last constraint and yields the following convex optimization problem:
\begin{subequations}
\begin{align}
\underset{X(\vect{x}), Y(\vect{x})}{\operatorname{maximize}} &~~~ \Real{E_{\mathrm{noncoro}, \lambda_0}(\vect{0})} \label{eq:final_optim:obj}\\
\operatorname{subject~to} &~~~ |E_{\mathrm{coro},\lambda}(\vect{k})|^2 < 10^{-c(\vect{k})} |E_{\mathrm{noncoro}, \lambda}(\vect{k})|^2\label{eq:final_optim:constr1}\\
\nonumber&~~~~~~~~~~~~~~~~~~~~~\forall~\vect{k} \in D~\forall~\lambda \in [\lambda_-, \lambda_+] \\
&~~~ X^2(\vect{x}) + Y^2(\vect{x}) \leq 1 \forall \vect{x}.\label{eq:final_optim:constr2}
\end{align}
\label{eq:final_convex_optim}
\end{subequations}
This problem can easily be solved using standard large-scale optimization algorithms, such as those implemented in Gurobi \citep{gurobi}. \changed{This} convexified problem does not guarantee \changed{a} phase-only solution, but we will see that in practice all solutions turn out to be phase only. \changed{Furthermore}, similarly to SPCs and APPs as mentioned above, the solutions space for APLCs is a subspace of this complex-amplitude apodizer optimization. As the latter produces PAPLCs in practice, a PAPLC will always \changed{perform the same or better than} an APLC for a given telescope pupil, dark zone geometry and design contrast.

\subsection{Symmetry considerations}
\label{sec:optimization:symmetry}

In general symmetric optimization problems are guaranteed to yield symmetric globally-optimal solutions if the optimization problem has multiple solutions \citep{waterhouse1983symmetric}. Applying the symmetry transformation to one globally-optimal solution can yield a different, but also globally-optimal solution. In our case, the final optimization problem is convex, and as such has only a single, unique solution, so any symmetry in the optimization problem must also be satisfied by the unique solution.

Making use of these symmetries can significantly reduce the computational complexity of the optimization. For example, for a point-symmetric focal-plane mask $M(\vect{k})=M(-\vect{k})$ and a point-symmetric dark zone ($-\vect{x}\in D~\forall~\vect{x} \in D$), the transformation $Y(\vect{x}) \to -Y(\vect{x})$ is a symmetry of the problem. Therefore $Y(\vect{x})=-Y(\vect{x})=0~\forall~\vect{x}$ and the complex transmission of the apodizer is real-valued. The optimization problem is now significantly simplified. The only remaining non-linear (in this case quadratic) constraint in Equation~\ref{eq:final_optim:constr2} can be replaced by two linear constraints. This yields a linear program, which is extremely easy to solve, even for a large number of variables.

Another interesting example is that of circular symmetry. If the telescope aperture, focal-plane mask, Lyot stop and dark zone are circularly symmetric, then the apodizer must consist of rings and must be completely real-valued (as circular symmetry implies point symmetry). This yields in practice an apodizer consisting of rings of zero and $\pi$ phase. This simplification significantly reduces the dimensionality of the solution space, thereby substantially reducing the computational complexity, which enables more extensive parameter studies, as shown in Section~\ref{sec:parameter_study:point-symmetric}.

\subsection{Tip-tilt correction for one-sided dark zones}
\label{sec:optimization:tiptilt}

For one-sided dark zones, the contrast is constrained only on one side of the PSF. In this case the optimizer tends to \changed{add a small tilt on the phase solution. The reason for this is that the optimizer maximizes the real part of the non-coronagraphic PSF at the optical axis, not at its peak. This seemingly tiny difference allows the optimizer to shift the peak of the non-coronagraphic PSF slightly in cases where the decrease in flux at the optical axis due to the shifted PSF is compensated by the increase in coronagraph throughput due to a less aggressive phase plate design. This centroid shift is unwanted as it effectively increases the inner working angle of the coronagraph. This effect is particularly prevalent for aggressive designs with small inner working angles, as a lot of throughput can be gained from shifting the PSF by a small amount. In these cases, the optimizer will produce a design with a larger inner working angle than what was asked.}

\changed{The same} effect is also commonly seen when optimizing one-sided APPs \citep{por2017optimal}, and we deal with it here in the same way. We constrain the intensity of the non-coronagraphic PSF to be smaller or equal to the intensity at the center of the non-coronagraphic PSF. This ensures that the maximum of the non-coronagraphic PSF is always attained at \changed{the optical axis so that any movement of the centroid of the planet is not allowed.} Mathematically, this constraint is expressed as
\begin{equation}
    |E_{\mathrm{noncoro},\lambda_0}(\vect{k})|^2 \leq |E_{\mathrm{noncoro}, \lambda_0}(\vect{0})|^2~~\forall~\vect{k}.
\end{equation}
This constraint is convex, and does therefore not affect convexity of the optimization problem. Despite this, the resulting optimization problem is in practice extremely slow to solve, due to the quadratic nature of the added constraint. Adopting a linearized version of this constraint, akin to \cite{por2017optimal}, yields an order of magnitude improvement in run time. A complete version of the optimization problem can be found in Appendix~\ref{app:full_optimization_problem}, including all approximations and modifications necessary to create an efficient numerical optimization problem.

\section{Parameter study for point-symmetric dark zones}
\label{sec:parameter_study:point-symmetric}

First we discuss point-symmetric dark zones. As this case is extremely similar to APLCs, we compare the PAPLC directly to the equivalent APLC. These APLCs are obtained using a similar optimization procedure. This can be derived starting from Equation~\ref{eq:final_convex_optim}, setting $Y(\vect{x})=0$ and additionally constraining $X(\vect{x}\geq 0$. This optimization problem for APLCs is equivalent to that used by \cite{zimmerman2016shaped}.

\begin{figure*}
    \plotone{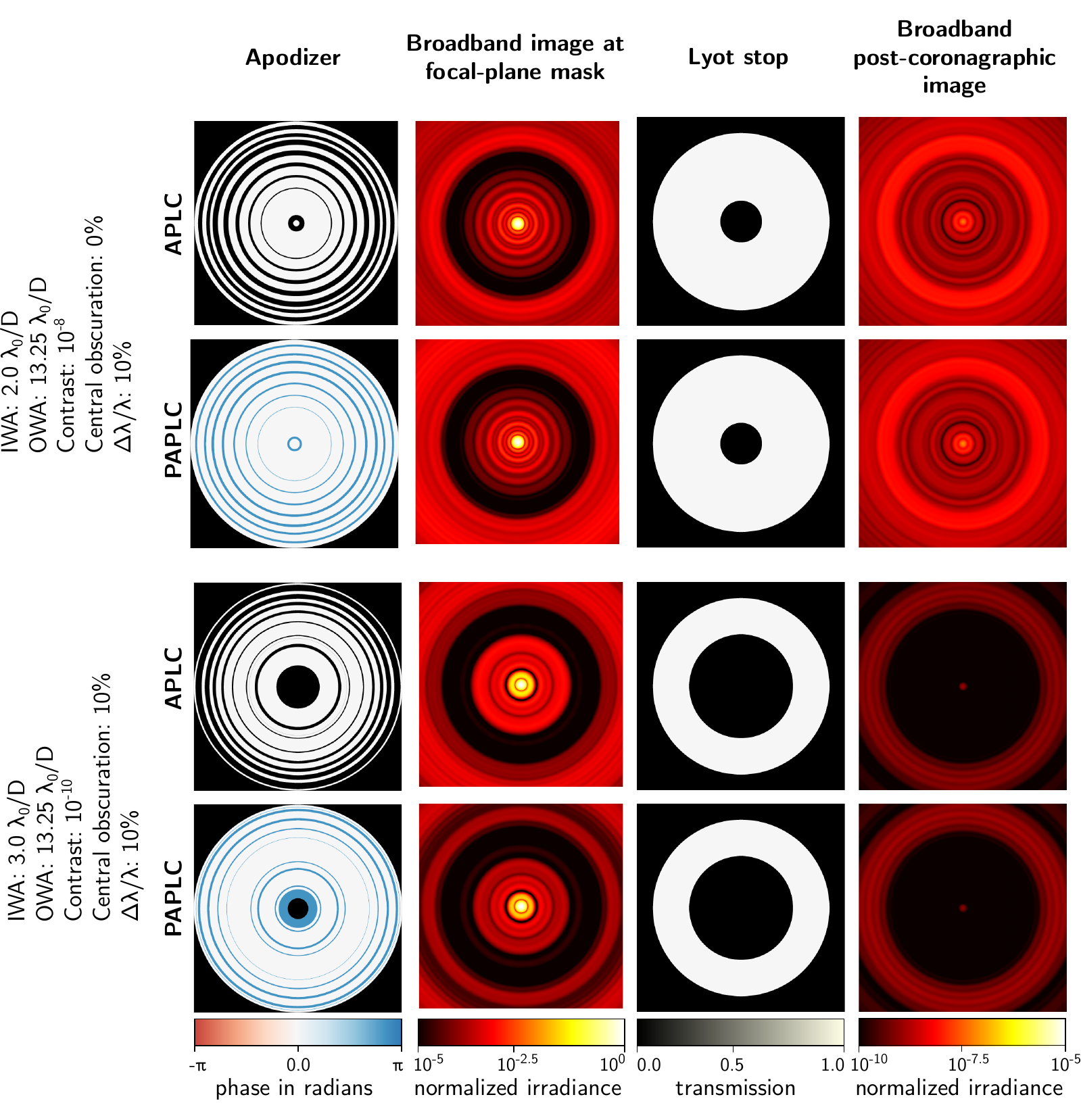}
    \caption{Some examples of PAPLC designs with point-symmetric dark zones. For two sets of parameters, we show both the APLC design and the PAPLC design. The phase patterns for the PAPLC consist of regions of $0$ or $\pi$ radians in phase, while the APLC designs consist of regions of $0$ and $1$ transmission. We show a 10\% broadband image just in front of the focal-plane mask in log-scale from $10^{-5}$ to $10^{0}$, and the post-coronagraphic image in log-scale from $10^{-10}$ to $10^{-5}$. The Lyot stop and focal-plane mask are optimized as hyper parameters.}
    \label{fig:examples_360}
\end{figure*}

To show the performance of a PAPLC, we use simplified telescope pupils. We use a circular telescope pupil with a circular central obscuration with a fractional size of $\mathit{CO} = D_\mathit{CO} / D_\mathit{tel}$. We use an annular Lyot mask parameterized by an inner and outer diameter, $L_\mathit{ID}$ and $L_\mathit{OD}$ respectively. These masks are shown schematically in Figure~\ref{fig:mask_definitions}. We will use an annular focal-plane mask, parameterized by an inner and outer diameter, $f_\mathit{ID}$ and $f_\mathit{OD}$ respectively. The dark zone is also annular, parameterized by an inner and outer radius $\mathit{DZ}_\mathit{min} \geq f_\mathit{ID}/2$ and $\mathit{DZ}_\mathit{max} \leq f_\mathit{OD}/2$. These masks are shown schematically in Figure~\ref{fig:mask_definitions}.

In Figure~\ref{fig:examples_360} we show some example PAPLCs along with equivalent APLC designs. Overall we can see that the ring structure in the PAPLCs is very similar to that of the APLCs. The rings are smaller by about a factor of two, which is to be expected as the apodization in phase has twice the effect of a zero transmission ring, however the rings are at the same position.

\begin{figure*}
    \plotone{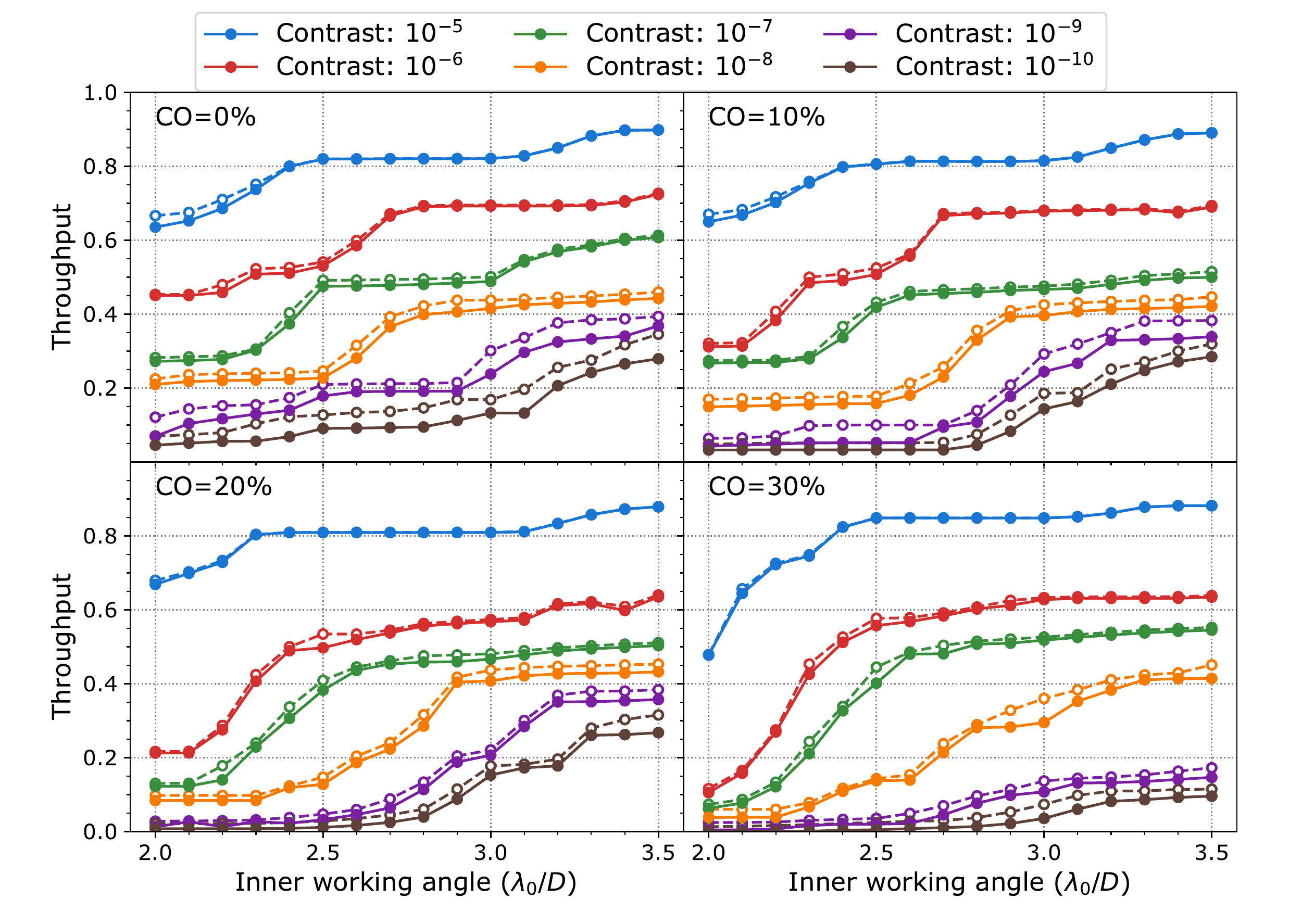}
    \caption{Throughput vs inner working angle for various contrasts for an annular dark zone. Solid lines and solid points are APLC designs, dashed lines and open points are PAPLC designs. The design contrast ranges from $10^{-5}$ to $10^{-10}$. Each point is a coronagraph design for which all hyperparameters (focal-plane mask size, and Lyot stop inner and outer diameters) have been optimized.}
    \label{fig:throughput_iwa_contrast_co_360}
\end{figure*}

We perform a full parameter study on the PAPLC and compared it to the similar APLC parameter study. We let the dark zone inner diameter change from $\mathit{DZ}_\mathit{min}=2.0 \lambda_0/D$ to $\mathit{DZ}_\mathit{min}=3.5 \lambda_0/D$, and fix the dark zone outer diameter at $\mathit{DZ}_\mathit{max}=13.25\lambda_0/D$. We vary the focal-plane mask inner diameter from $f_\mathit{ID}=2\mathit{DZ}_\mathit{min}-5\lambda_0/D$ to $f_\mathit{ID}=2\mathit{DZ}_\mathit{min}$. The focal-plane mask outer diameter is fixed at $f_\mathit{OD} = 2\mathit{DZ}_\mathit{max}$, as it was found to have no influence on the throughput of both the PAPLC and the APLC. We vary the Lyot mask inner diameter from $L_\mathit{ID}=CO$ to $L_\mathit{ID}=CO + 0.4$, and the outer diameter from $L_\mathit{OD}=0.85$ to $L_\mathit{OD}=1$. The relative spectral bandwidth was $10\%$. We performed the parameter study for design contrasts from $10^{-5}$ to $10^{-10}$ with central obscuration ratios varying from $0\%$ to $30\%$, to represent a full range of potential ground-based and space-based instrument parameters.

In Figure~\ref{fig:throughput_iwa_contrast_co_360} we show the maximum throughput for a combination of dark zone inner diameter, central obscuration ratio and design contrast, where all other hyperparameters have been optimized out using the brute-force optimization procedure in Section~\ref{sec:optimization:problem}. APLCs are denoted by filled points and solid lines, while the PAPLC has open points and dashed lines. It is clear that PAPLCs for point-symmetric dark zones do not hold a big advantage over APLCs. Only when throughput is already compromised, the PAPLC can gain a significant advantage, at most $\sim50\%$ in this parameter space. 

Also clear is the plateau behaviour of the throughput: at some points the throughput can be almost insensitive to dark zone inner diameter, while at other points the throughput can drop rapidly for even a small change in dark zone inner diameter. This drop in throughput occurs every $0.5$ to $1\lambda_0/D$. The drops change their center position as function of central obscuration ratio and contrast, and can sometimes merge. This behaviour is similar to that of APPs and shaped pupils with annular dark zones \cite{por2017optimal}.

In conclusion: the PAPLC is marginally better than the APLC, but the difference between them is extremely minor, easily overshadowed by the ease of manufacturing of binary amplitude masks. Only where the throughput is low, the PAPLC offers a large relative, but small absolute, performance gain.

\section{Parameter study for one-sided dark zones}
\label{sec:parameter_study:one-sided}

\begin{figure*}
    \plotone{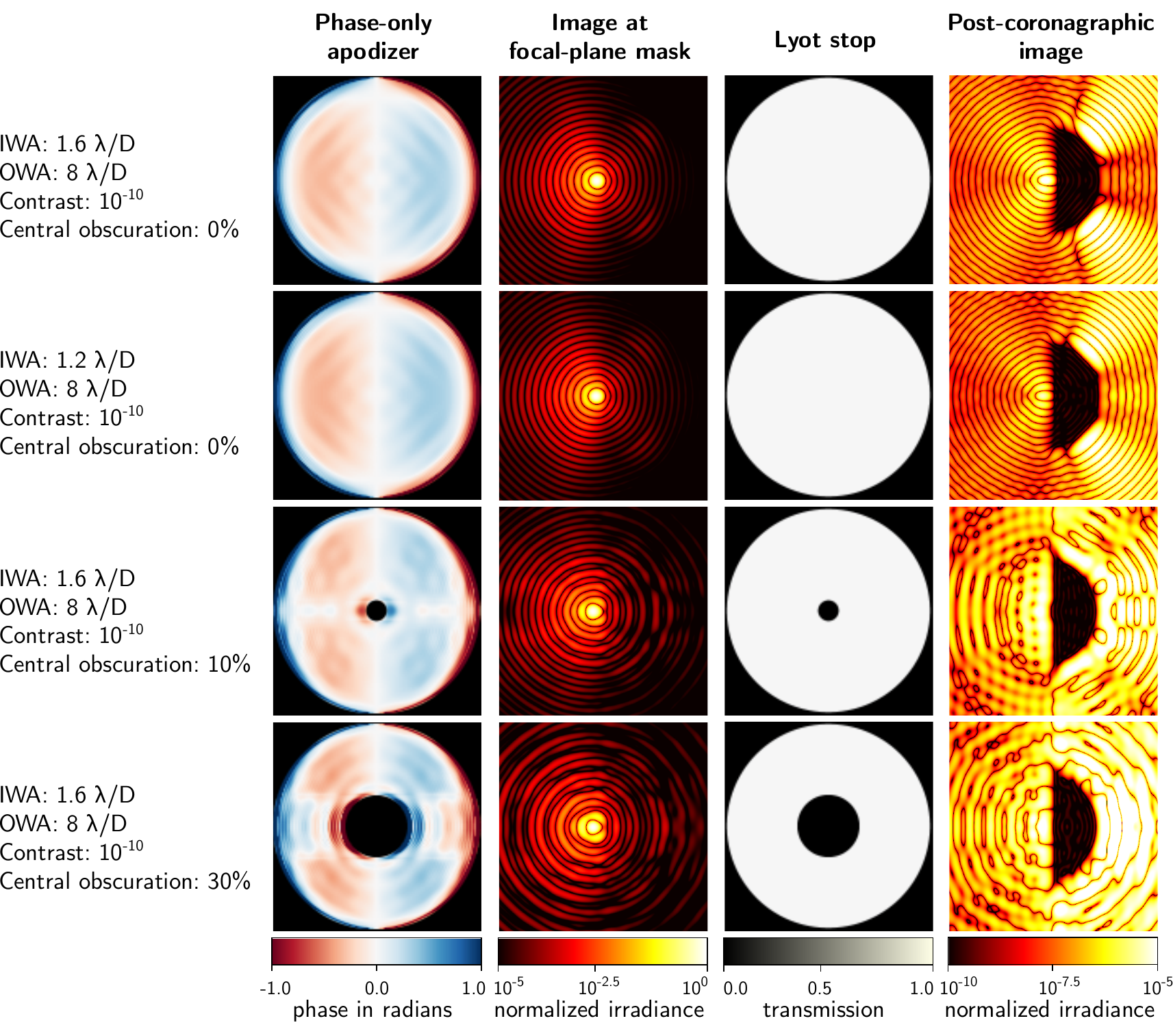}
    \caption{Some examples of PAPLC designs with one-sided dark zones. The color scale for phase is from $-1 \mathrm{~rad}$ to $1 \mathrm{~rad}$ but typically the phase pattern rms is $\sim0.4\mathrm{~rad}$. We show the image at the focal-plane mask with a translucent focal-plane mask to show the positioning of the focal-plane mask relative to the peak of the PSF. In the coronagraph the focal-plane mask is completely opaque. The image at the focal-plane mask is in log-scale from $10^{-5}$ to $10^0$. The post-coronagraphic is also in log scale from $10^{-10}$ to $10^{-5}$. \changed{The focal-plane mask offset, and Lyot-stop inner and outer diameters were optimized to maximize post-coronagraphic throughput.}}
    \label{fig:examples_180}
\end{figure*}

\begin{figure*}
    \plotone{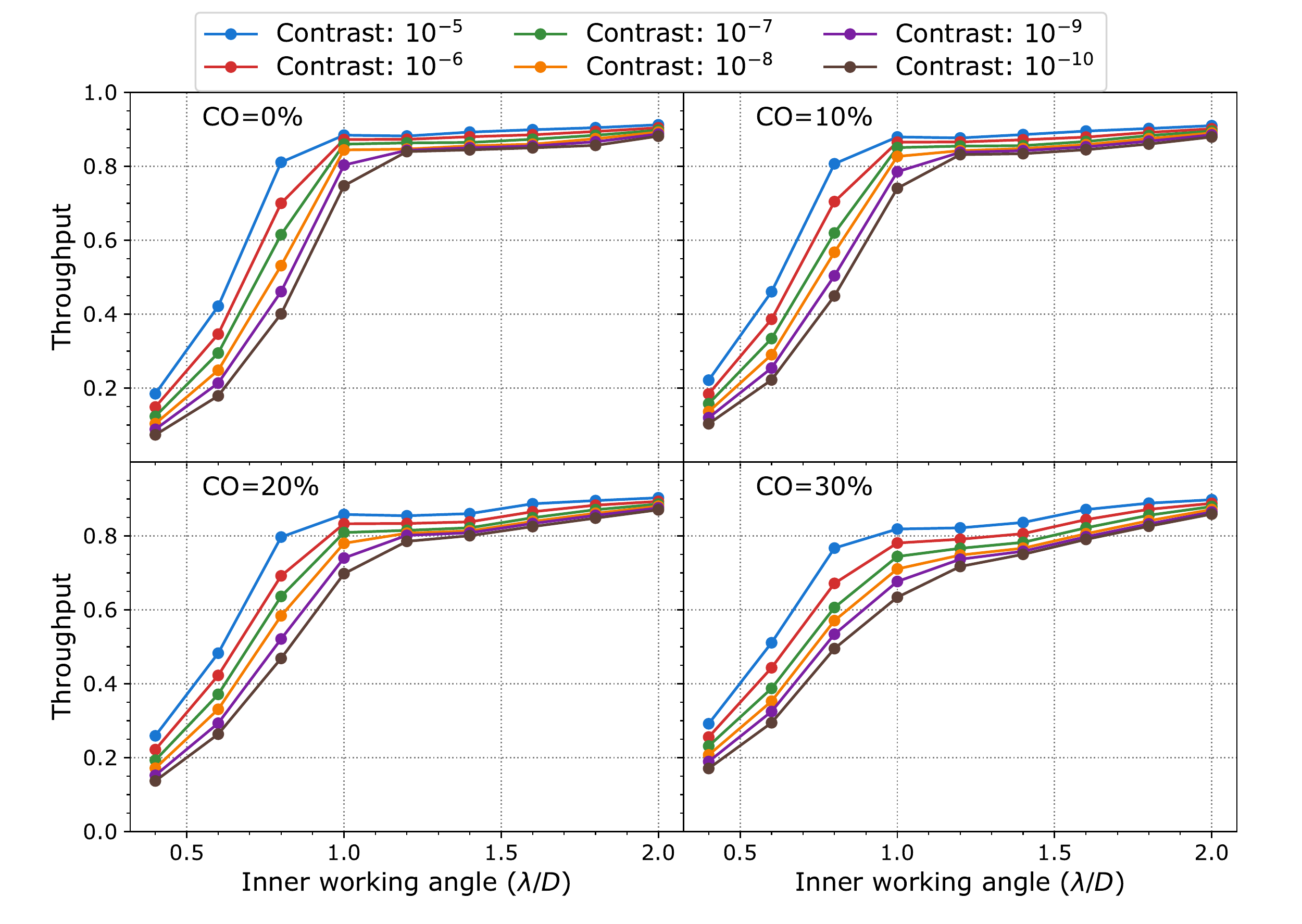}
    \caption{Throughput vs inner working angle for various contrasts for a one-sided dark zone. All designs are PAPLC designs. The design contrast ranges from $10^{-5}$ to $10^{-10}$. Each point is a coronagraph design for which all hyperparameters (focal-plane mask offset, and Lyot stop inner and outer diameters) have been optimized. \changed{Each of the example designs in Figure~\ref{fig:examples_180} correspond to a point in this figure.}}
    \label{fig:throughput_iwa_contrast_co_180}
\end{figure*}

As phase-only apodizers can bring about one-sided dark zones, it is interesting to look at a Lyot-style coronagraph based on a one-sided dark zone. We use a focal-plane mask that blocks all the light on one side of the focal-plane. This mask is offset from the center of the PSF by $f_\mathit{edge}$. We again use an annular Lyot stop. The dark zone is D-shaped on the side of the PSF that is not blocked by the focal-plane mask. These masks are shown schematically in Figure~\ref{fig:mask_definitions}.

\changed{The propagation through the focal-plane mask is performed using standard forward and backward FFTs on a zero-padded pupil. As the knife-edge is invariant across the y-axis, we can view all rows of the pupil as independent and avoid performing an FFT across the y-axis, as well as all FFTs across the x-axis on the zero-padded rows. This makes for a much faster propagation and reduced memory usage. An implementation of this method is available in the open-source package HCIPy~\citep{por2018hcipy}.}

We show some examples in Figure~\ref{fig:examples_180}. We can see that the phase apodizer acts as an APP, in that it creates a one-sided dark zone with a deepening raw contrast as function of angular separation. At no point however does the stellar PSF at the focal-plane mask reach the required design contrast. The design raw contrast is produced by the focal-plane mask and the Lyot-stop mask, deepening the contrast by more that three decades.

\subsection{Contrast, inner working angle and central obscuration ratio}

We perform a full parameter study on the PAPLC for one-sided dark zones. We let the dark zone inner radius change from $\mathit{DZ}_\mathit{min}=0.4\lambda_0/D$ to $\mathit{DZ}_\mathit{min}=2.0\lambda_0/D$, and fixed the outer radius at $\mathit{DZ}_\mathit{max}=8\lambda_0/D$, mainly limited by the computational run time for the full parameter study. We varied the focal-plane mask offset from $f_\mathit{edge}=\mathit{DZ}_\mathit{min}$ to $f_\mathit{edge}=\mathit{DZ}_\mathit{min} - 1.0 \lambda_0/D$. The Lyot-mask parameters are varied in the same way as for the point-symmetric dark zone. All masks were calculated for a single wavelength only: we presume monochromatic light. We performed the parameter study for design contrasts from $10^{-5}$ to $10^{-10}$ with central obscuration ratios varying from $0\%$ to $30\%$, to represent a full range of potential ground-based and space-based instrument requirements.

In Figure~\ref{fig:throughput_iwa_contrast_co_180} we show the maximum throughput for a combination of dark zone inner diameter, central obscuration ratio and design contrast, where all other parameters have been optimized out. Shrinking the Lyot stop had no positive effects on the throughputs: having the Lyot stop the same as the telescope pupil yielded the best throughput. Also clear is that for dark zone inner radii of $\gtrapprox 1.2\lambda_0/D$ the throughput is relatively independent of design contrast. This is a useful property for coronagraphs destined for space-based instruments. We also see that throughput at a fixed dark zone inner radius is relatively insensitive to central obscuration ratio of the telescope pupil.

\subsection{Achromatization and residual atmospheric dispersion}

We can produce an achromatic design from any monochromatic design by centering the focal-plane mask (ie. using $f_\mathrm{edge}=0$) and introducing a wavelength-dependent shift using a phase tilt at the phase-only apodizer. \changed{This phase tilt acts in the same way as the phase pattern, so we can simply modify the apodizer pattern by adding a tilt on it.} In this way, as the PSF grows with wavelength, it will offset the PSF by the same amount, leaving the edge of the focal-plane mask in the same position relative to the rescaled PSF. This makes the one-sided PAPLC completely achromatic in theory (barring experimental effects). One possible downside to this practice is that the planetary PSF inherits this phase tilt, which acts as a grating smearing out its light across the detector. For small focal-plane mask offsets however, this effect can be quite small. For example, for a relative spectral bandwidth of $\Delta\lambda/\lambda_0=20\%$, and a focal-plane offset of $f_\mathit{edge}=1.6\lambda/D$, the planet is smeared out across $\Delta\lambda/\lambda_0 \cdot f_\mathit{edge} = 0.32\lambda_0/D$, well within the size of the Airy core of the planet. This smearing is independent of field position.

The focal-plane mask is translation invariant in one direction. This means that any tip-tilt errors in that direction will have no influence on the coronagraphic performance other than movement of the coronagraphic PSF. We will explore the tip-tilt senstivity of the PAPLC further in Section~\ref{sec:case_studies:performance}. Here we focus on the application of this insensitivity for residual atmospheric dispersion for ground-based telescopes. As telescopes get larger, atmospheric dispersion will become stronger relative to the size of the Airy core, making the performance of the atmospheric dispersion corrector even more critical for future large ground-based telescopes \citep{pathak2016high}.

As the PAPLC is insensitive to tip-tilt along one axis, we can align the residual atmospheric dispersion along the knife edge. In this case, the atmospheric dispersion doesn't degrade the coronagraph performance, and we would only require $\lesssim 1~\lambda_0/D$ of residual atmospheric dispersion, instead of less than a few \changed{tenths} to \changed{hundredths} of $\lambda_0/D$ for other focal-plane coronagraphs. This significantly relaxes the constraints on the atmospheric dispersion correctors and simplifies their implementation and complexity. Of course, this is only possible on telescopes where the orientation of the pupil is fixed with respect to the zenith, which is the case for all alt-azimuth-mounted telescopes, the majority of current large telescopes.

\begin{figure}
    \plotone{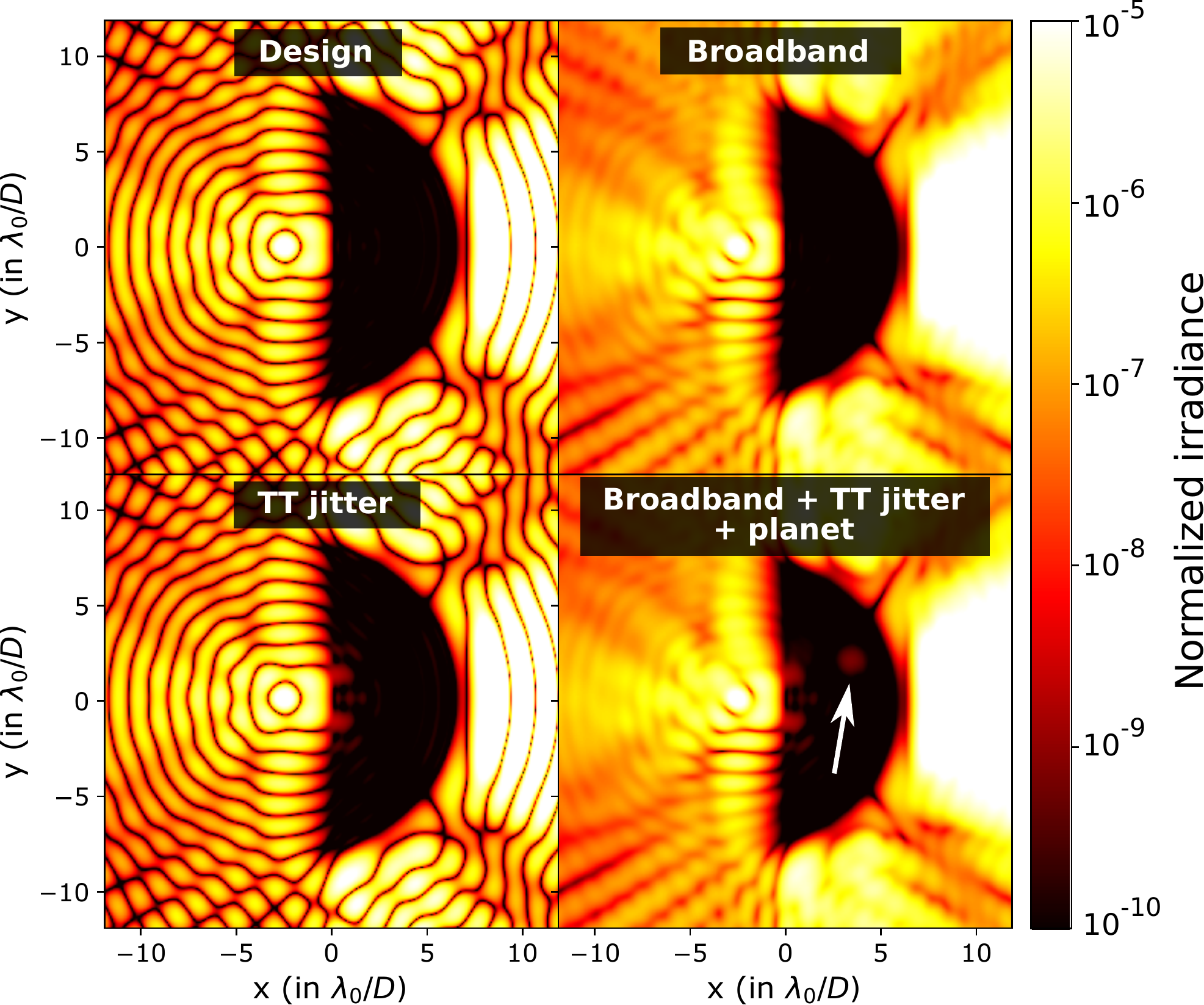}
    \caption{Raw post-coronagraphic images for a one-sided dark zone with an inner working angle of $1.6\lambda/D$ with increasing imperfections. \emph{Top left:} Only tip-tilt jitter with $0.003\lambda/D$ rms. \emph{Top right:} tip-tilt jitter and $20\%$ broadband light. \emph{Bottom left:} tip-tilt jitter, broadband light and $0.5 \lambda_0/D$ residual dispersion from the ADC. \emph{Bottom right:} tip-tilt jitter, broadband light, residual dispersion and a planet, indicated with an arrow, with a raw contrast of $10^{-9}$ relative to the host star.}
    \label{fig:chromaticity}
\end{figure}

In Figure~\ref{fig:chromaticity} we show each of these effects for an example PAPLC design. We show the design PAPLC post-coronagraphic PSF, a post-coronagraphic PSF with (isotropic) tip-tilt jitter and a broadband light source, a post-coronagraphic PSF with broadband light and a $0.5\lambda/D$ residual atmospheric dispersion pointed along the focal-plane mask edge, and finally a post-coronagraphic PSF with (isotropic) tip-tilt jitter, residual atmospheric dispersion, broadband light and an injected planet.

\section{Case studies for VLT/SPHERE and LUVOIR-A}
\label{sec:case_studies}

To show that the PAPLC can handle more complicated apertures as well, we present two case studies. The first is a design for VLT/SPHERE, showing that the design method can deal with a complex telescope pupil consisting of spiders and dead \changed{deformable mirror} actuators. The second is a design for LUVOIR-A, showing that designs with space-based contrasts are possible, and showing that the PAPLC can handle the segmented telescope pupil with spiders and central obscuration seen in future large space telescopes.

\begin{figure*}
    \plotone{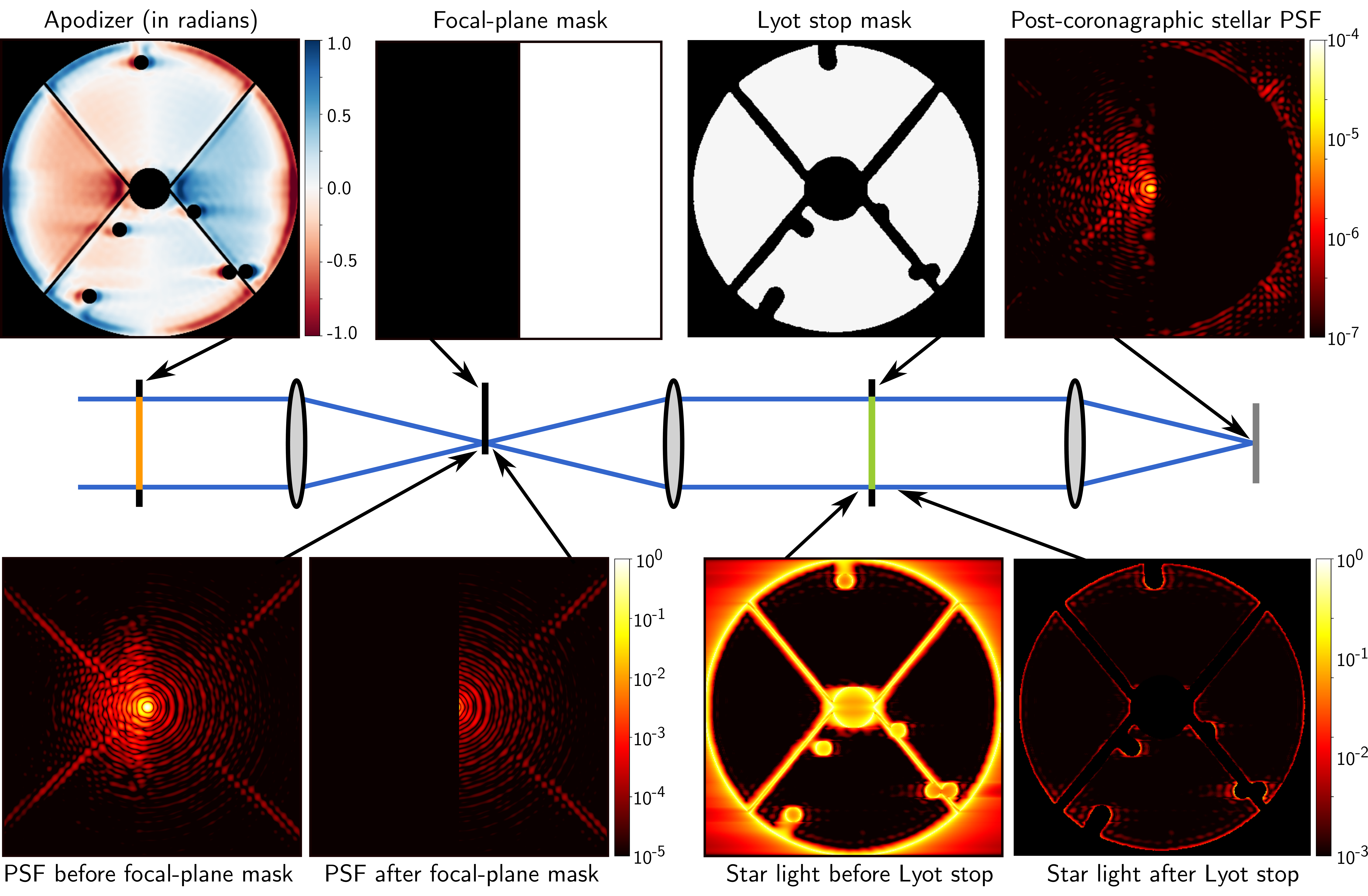}
    \caption{The case study design for VLT/SPHERE. We show the apodizer phase pattern, focal-plane mask and Lyot stop. Additionally, we show the light in each of the coronagraphic planes: before and after the focal-plane mask (on a logarithmic scale), and before and after the Lyot stop (on a logarithmic scale, normalized to the peak intensity). Finally, the normalized irradiance of the post-coronagraphic stellar PSF is shown (on a logarithmic scale). Note that the peak in the post-coronagraphic stellar PSF is not the Airy core, but rather a stellar leakage at a relative intensity of $\sim 2\times10^{-4}$ that of the star PSF.}
    \label{fig:case_study:sphere:design}
\end{figure*}

\subsection{VLT/SPHERE}
\label{sec:case_studies:sphere}

\changed{As VLT/SPHERE is a ground-based instrument, it contains an AO system that will limit the raw contrast of resulting images to a level of $\sim10^{-4}$ to $\sim10^{-6}$. We fix the design raw contrast at $10^{-7}$ to avoid having the coronagraph limit the raw contrast of observations.} The outer working angle was fixed at $30\lambda/D$. For the Lyot mask we used that of the existing ALC2 Lyot mask in VLT/SPHERE \citep{guerri2011apodized} to simplify integration in the VLT/SPHERE instrument. We performed a small parameter study on the inner working angle, of which we present here only one of the solutions. This solution has an inner working angle of $1.4\lambda/D$ and a focal-plane mask offset of $f_\mathit{edge}=1.0\lambda/D$. We show the phase solution, PSF on the focal-plane mask, intensity at the Lyot stop and post-coronagraphic PSF in Figure~\ref{fig:case_study:sphere:design}.

\begin{figure*}
    \plotone{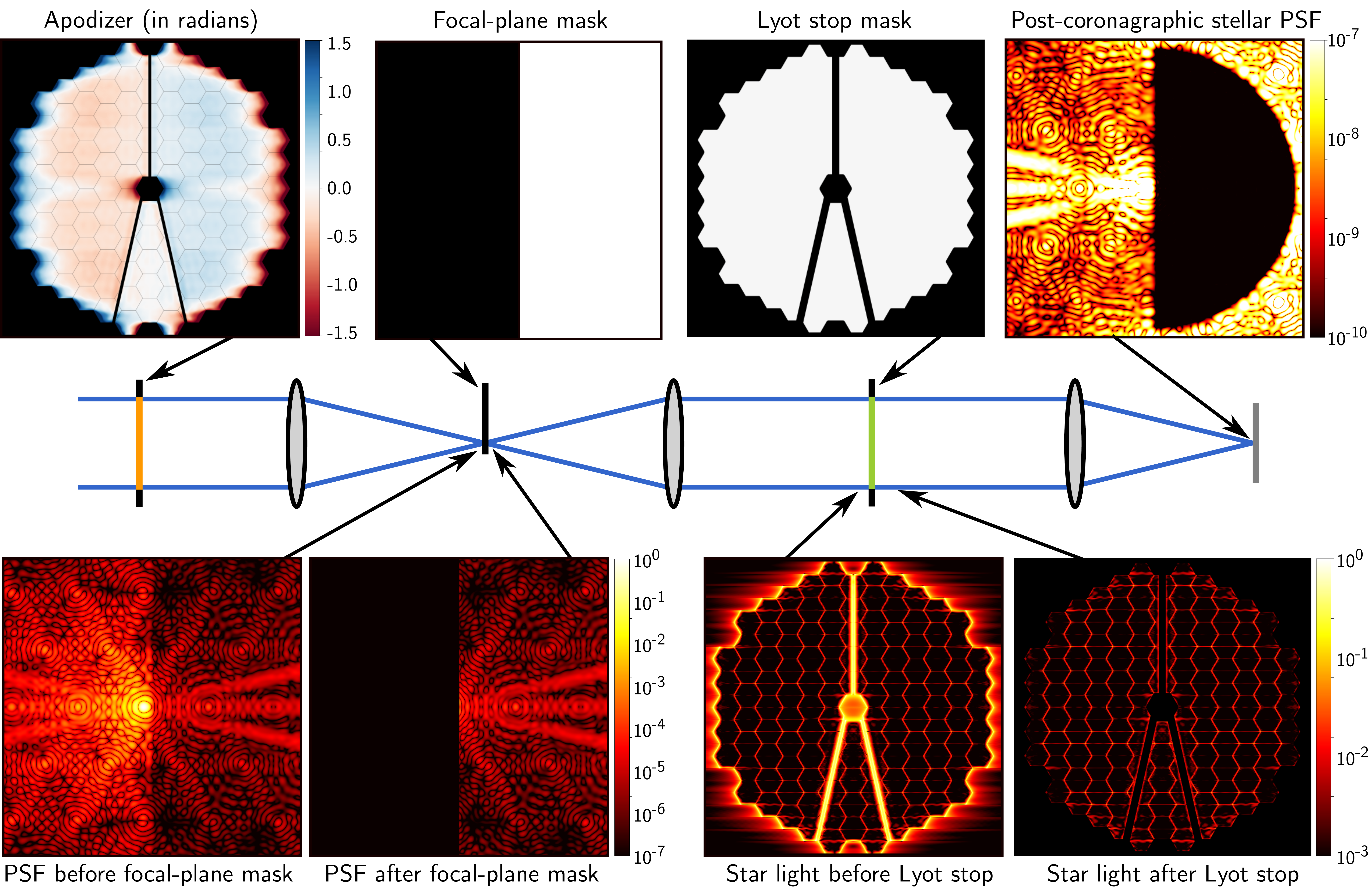}
    \caption{The case study design for the LUVOIR-A telescope. We show the apodizer phase pattern, focal-plane mask and Lyot stop. Additionally, we show the light in each of the coronagraphic planes: before and after the focal-plane mask (on a logarithmic scale), and before and after the Lyot stop (on a logarithmic scale, normalized to the peak intensity). Finally, the normalized irradiance of the post-coronagraphic stellar PSF is shown (on a logarithmic scale). Note that the peak in the post-coronagraphic stellar PSF is not the Airy core, but rather a stellar leakage at a relative intensity of $\sim 2\times10^{-5}$ that of the star PSF.}
    \label{fig:case_study:luvoir:design}
\end{figure*}

The light at the positions of the dead actuators on the deformable mirror in VLT/SPHERE are blocked at the apodizer. This provides greater resilience against the unknown positions of the dead actuators. For traditional Lyot coronagraphs and also APLCs, dead \changed{deformable mirror} actuators are usually blocked in the Lyot stop. This however requires a small blocking element in the focal-plane mask, as in this case the local perturbation caused by the dead \changed{deformable mirror} actuator is kept local by the focal-plane mask making it possible to efficiently block its resulting speckles in the Lyot stop. In our case however, the focal-plane mask blocks \changed{over} half of the field of view, making it necessary for the light impinging on dead actuators \changed{on the deformable mirror} to be blocked upstream at the apodizer, as speckles caused by \changed{a} dead actuator are now spread out in the Lyot stop. Also the support structure of the secondary mirror has been thickened, the secondary obscuration broadened and the outer diameter of the pupil shrunk to accommodate a misalignment \changed{in translation} of the apodizer \changed{of up to $0.5\%$ of the diameter of the re-imaged telescope pupil}.

\subsection{LUVOIR-A}
\label{sec:case_studies:luvoir}

As LUVOIR-A is a space telescope, we fix the design raw contrast at $10^{-10}$. The outer working angle was also fixed at $30\lambda/D$. For the Lyot mask we used a thickened version of the LUVOIR-A pupil, where segment gaps, spiders and central obscuration were broadened and the outer diameter was shrunk by $\sim1.5\%$. No attempt was made to optimize this percentage as a hyperparameter. We performed a small parameter study on the inner working angle, of which we present here only one of the solutions. This solution has an inner working angle of $2.2\lambda/D$ and a focal-plane mask offset of $f_\mathit{edge}=1.8\lambda/D$. We show the phase solution, PSF on the focal-plane mask, intensity at the Lyot stop and post-coronagraphic PSF in Figure~\ref{fig:case_study:luvoir:design}.

\subsection{Performance}
\label{sec:case_studies:performance}

\begin{figure}
    \plotone{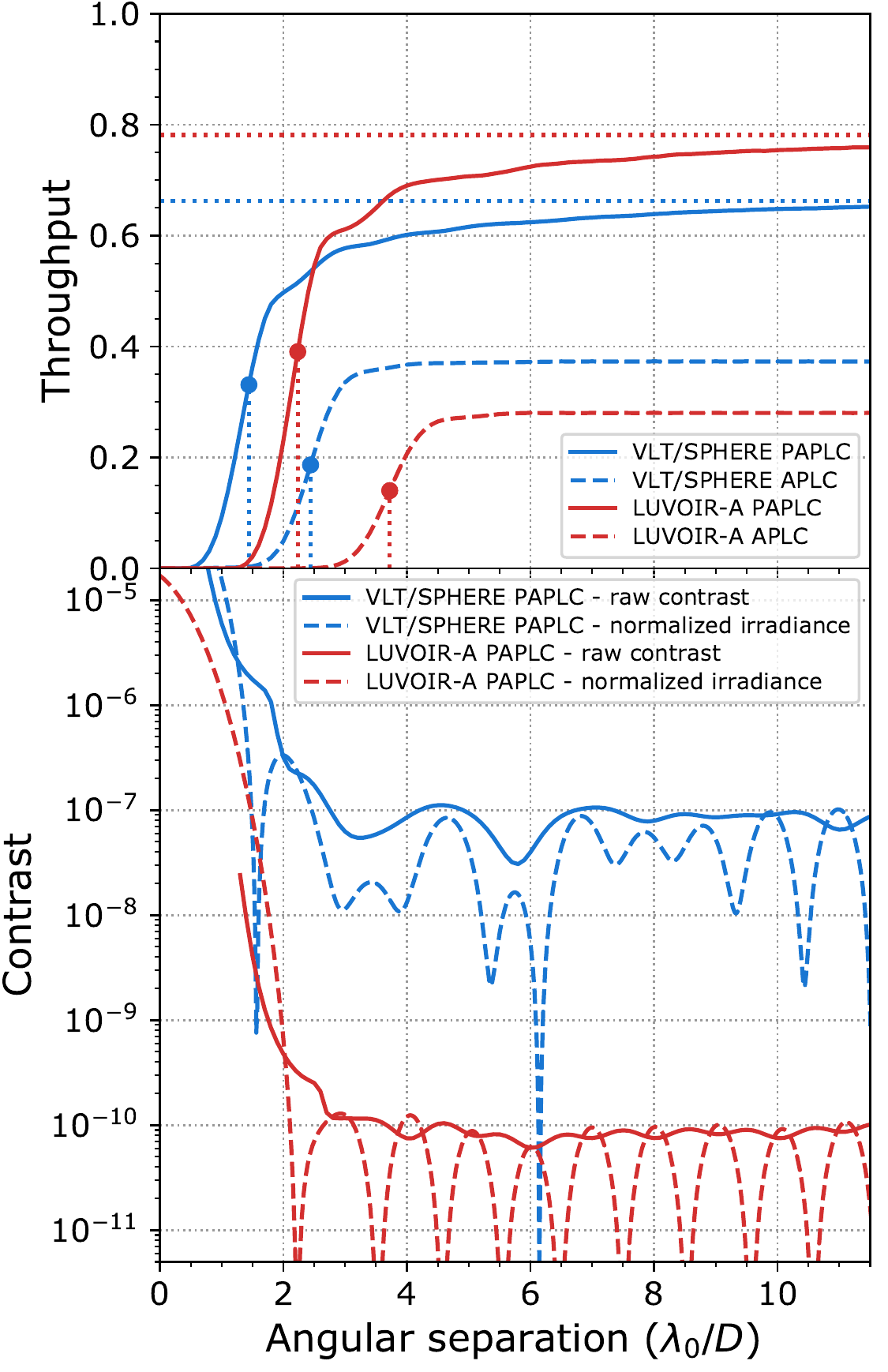}
    \caption{The throughput, raw contrast and normalized irradiance for both the VLT/SPHERE and LUVOIR-A designs. Also shown are the throughput for APLC designs for each telescope. Note that the PAPLC has \changed{a smaller} field of view compared to the APLC designs, which should be taken into account during survey mode but is irrelevant in characterization mode. The inner working angles and maximum throughput for each of the coronagraph designs are listed in Table~\ref{tab:throughput_iwa_aplc_paplc}.}
    \label{fig:case_study:both:contrast_throughput}
\end{figure}

We show the throughput and contrast for both case studies in Figure~\ref{fig:case_study:both:contrast_throughput}. We see that the inner working angles for the two coronagraph designs is $1.4\lambda/D$ for VLT/SPHERE and $2.2\lambda/D$ for LUVOIR-A. At larger angular separations the throughput rises quickly, reaching $90\%$ of its maximum throughput at $4\lambda/D$ and $4.2\lambda/D$ for the VLT/SPHERE and LUVOIR-A design respectively.

The maximum throughput is $66\%$ and $78\%$ for the VLT/SPHERE and LUVOIR-A design respectively. For the VLT/SPHERE design this maximum throughput is primarily limited by the Lyot mask. The throughput without phase-apodizer is $\sim69\%$, and the addition of any phase pattern on top can only reduce the throughput from there on. The throughput for the LUVOIR-A design however is shared between the phase apodization and the Lyot stop: without the Lyot-stop the throughput is $\sim87\%$.

We also show the throughput for novel APLC designs for the VLT/SPHERE instrument and LUVOIR-A telescope. The VLT/SPHERE APLC design is a preliminary solution for a possible future upgrade of VLT/SPHERE (courtesy Mamadou N'Diaye). The LUVOIR-A APLC design is a part of a coronagraph design study for the LUVOIR-A aperture (courtesy R\'{e}mi Soummer). Their design procedure for both is based on the hybrid shaped pupil/APLC designs by \cite{ndiaye2016apodized}. The inner working angle and maximum throughput of the PAPLC and APLC designs are summarized in Table~\ref{tab:throughput_iwa_aplc_paplc}. Care must be taken when directly comparing throughput between APLC and PAPLC designs, due to their different field of views. During survey mode, one needs to \changed{observe at several sky-rotation angles or roll angles to retrieve a complete image for the full field of view}, effectively reducing the throughput by \changed{a factor corresponding to the number of observations}. During characterization mode however, field of view is irrelevant, and a direct comparison can be made. The PAPLC designs yield almost double or triple the maximum throughput, for the VLT/SPHERE and LUVOIR-A design respectively, mostly or completely neutralizing the disadvantage in field of view. Furthermore, it provides a significantly reduced inner working angle by $1.0\lambda_0/D$ and $1.5\lambda_0/D$ for the VLT/SPHERE and LUVOIR-A designs respectively.

\begin{table}
\centering
\begin{tabular}{l|cc|cc}
& \multicolumn{2}{c}{VLT/SPHERE} & \multicolumn{2}{c}{LUVOIR-A} \\
Quantity & PAPLC & APLC & PAPLC & APLC \\
\hline
IWA & $1.4\lambda_0/D$ & $2.4\lambda_0/D$ & $2.2\lambda_0/D$ & $3.7\lambda_0/D$ \\
$T_\mathrm{max}$ & $66\%$ & $38\%$ & $78\%$ & $28\%$
\end{tabular}
\caption{The inner working angle and throughput for all coronagraph designs shown in Figure~\ref{fig:case_study:both:contrast_throughput}. Care must be taken when directly comparing maximum throughput between PAPLC and APLC designs, due to their different field of view. A discussion of theses quantities can be found in the text.}
\label{tab:throughput_iwa_aplc_paplc}
\end{table}

\begin{figure}
    \plotone{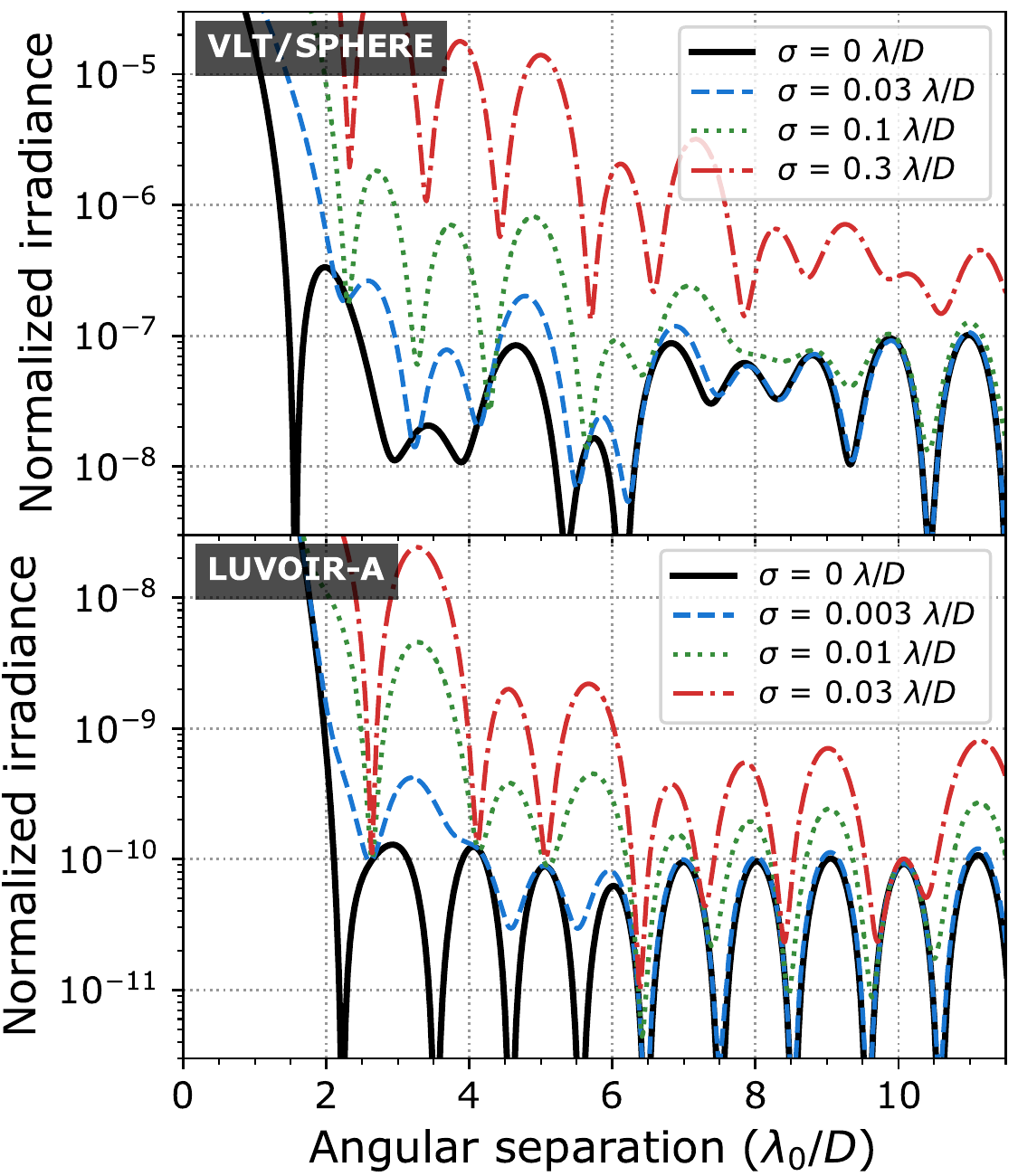}
    \caption{Slices of the normalized irradiance for varying values of the RMS tip-tilt error on the star for both the VLT/SPHERE and LUVOIR-A design. The different RMS values were chosen to show the transition from no effect to a significant effect on the normalized irradiance. A normal, isotropic distribution was assumed for tip-tilt.}
    \label{fig:case_study:both:tip_tilt_slices}
\end{figure}

To test the coronagraph as function of tip-tilt jitter of the on-axis source, we show slices of the normalized intensity at various values for tip-tilt errors in Figure~\ref{fig:case_study:both:tip_tilt_slices}. We assume a normal, isotropic distribution of the tip-tilt offset with a standard deviation of $\sigma$. For the VLT/SPHERE design a $<3\times10^{-6}$ contrast for angular separations $>2.1\lambda/D$ is still achieved with a tip-tilt rms of $\sigma<0.1\lambda/D$. This tip-tilt performance is (almost) achieved with current high-contrast imagers from the ground at infrared wavelengths \citep{fusco2014final,escarate2018vibration}. For the LUVOIR-A design, a contrast of $<5\times10^{-9}$ for angular separations $>2.5\lambda/D$ is achieved at a tip-tilt rms of $\sigma<0.01\lambda/D$. \changed{This tip-tilt sensitivity is significantly worse than the APLC for LUVOIR-A, and has to be improved for the PAPLC to be considered a viable option for giant space telescopes.}

Both designs presented in this section, in fact all designs presented in this work, are not made robust against aberrations or misalignment of the Lyot stop. As APLCs can be made robust to aberrations by including these aberrations in the optimization problem \citep{ndiaye2015apodized}, one can postulate that PAPLCs might be able to be made robust as well. The design of robust PAPLCs and an analysis of the corresponding hit in coronagraphic throughput is left for future work.

\section{Conclusions}
\label{sec:conclusions}

In this work we presented the phase-apodized-pupil Lyot coronagraph. This coronagraph uses a standard Lyot-style architecture and its design procedure is a mix between that for the APLC and the APP coronagraph. \changed{Starting from an aperture-photometric methodology, we derive a tractable optimization problem to obtain a globally-optimal solution for the phase pattern in the PAPLC. This shows} that an PAPLC will always \changed{perform equally or better} an APLC by design, \changed{given a certain focal-plane mask and Lyot-stop}, barring experimental or manufacturing errors.

We distinguished two cases for a PAPLC. The first uses a conventional annular focal-plane mask and produces point-symmetric dark zones. This case provides performance analogous to the APLC, showing similar structure in the apodizer design. Apodizers consist of regions of $0$ or $\pi$ radians in phase, rather than $0$ or $1$ in amplitude for the APLC.

The second case uses a knife-edge focal-plane mask and is optimized to produce a one-sided dark zone. This case yields apodizers similar to APPs, but use the Lyot stop to gain in contrast. These designs show inner working \changed{angles as close as $1.4\lambda/D$} and can be made entirely achromatic. \changed{Additionally the coronagraph can reach space-based contrasts ($<10^{-10}$) at these inner working angles at a throughput of around $60\%-80\%$ for central obscurations up to $30\%$.} Furthermore, as the knife edge is invariant to translation along one axis, the coronagraph can handle tilt along that axis as well. We can use this to make the coronagraph invariant to residual atmospheric dispersion.

We presented two designs for realistic telescope pupils: one for VLT/SPHERE as an example of a ground-based telescope, and one for LUVOIR-A as an example of a space-based telescope. This shows that the PAPLC can deal with blocking dead \changed{deformable mirror} actuators, secondary support structure and the segmentation in these telescope pupils.

Future research will focus on testing PAPLC in a lab setting and finally on sky. Additionally, making the PAPLC robust against low-order aberrations is certainly intriguing from a design perspective. Another interesting avenue for future research is integrating the PAPLC with wavefront sensing. As the light from the bright side of the PSF is blocked by the focal plane mask, one can envision using a reflective focal-plane mask instead, and reimaging the bright side on a separate, fast detector. Adding a defocus to this reimaged PSF allows reconstruction of the phase of the incoming wavefront using phase diversity \citep{Gonsalves1982} or spatial linear dark field control \citep{miller2017spatial}.

\section*{Funding}

EHP acknowledges funding by The Netherlands Organisation for Scientific Research (NWO) and the S\~{a}o Paulo Research Foundation (FAPESP).

\section*{Acknowledgments}

I thank Matthew Kenworthy and Christoph Keller for their comments, which helped improve this work. I also thank Mamadou N'Diaye, R\'{e}mi Soummer, Alexis Carlotti, R\'{e}mi Flamary, Kathryn St. Laurent and Jamie Noss for supplying the APLC designs for VLT/SPHERE and LUVOIR-A.

This research made use of HCIPy, an open-source object-oriented framework written in Python for performing end-to-end simulations of high-contrast imaging instruments \citep{por2018hcipy}. Additionally we used the numerical library NumPy \citep{walt2011numpy} and visualizations were made using the library Matplotlib \citep{hunter2007matplotlib}.

\section*{Disclosures}
The author declares that there are no conflicts of interest related to this article.

\appendix

\section{Full optimization problem}
\label{app:full_optimization_problem}

Here we state the full optimization problem, as solved by the large-scale optimization software. This includes linearized constraints on the contrast, and a linearized version of the tip-tilt correction algorithm as presented in Section~\ref{sec:optimization:tiptilt}.

\begin{subequations}
\begin{align}
\underset{X(\vect{x}), Y(\vect{x})}{\operatorname{maximize}} &~~~ \Real{E_{\mathrm{noncoro}, \lambda_0}(\vect{0})}\\
\operatorname{subject~to} &~~~ X^2(\vect{x}) + Y^2(\vect{x}) \leq 1 ~~\forall~\vect{x}\\
&~~~ \Real{E_{\mathrm{coro},\lambda}} + \Imag{E_{\mathrm{coro},\lambda}} \leq \sqrt{10^{-c(\vect{k})} S_\mathrm{expected}} ~~\forall~\vect{k} \in D~\forall~\lambda \in [\lambda_-, \lambda_+] \\
&~~~ \Real{E_{\mathrm{coro},\lambda}} - \Imag{E_{\mathrm{coro},\lambda}} \leq \sqrt{10^{-c(\vect{k})} S_\mathrm{expected}} ~~\forall~\vect{k} \in D~\forall~\lambda \in [\lambda_-, \lambda_+] \\
&~~~ -\Real{E_{\mathrm{coro},\lambda}} + \Imag{E_{\mathrm{coro},\lambda}} \leq \sqrt{10^{-c(\vect{k})} S_\mathrm{expected}} ~~\forall~\vect{k} \in D~\forall~\lambda \in [\lambda_-, \lambda_+] \\
&~~~ -\Real{E_{\mathrm{coro},\lambda}} - \Imag{E_{\mathrm{coro},\lambda}} \leq \sqrt{10^{-c(\vect{k})} S_\mathrm{expected}} ~~\forall~\vect{k} \in D~\forall~\lambda \in [\lambda_-, \lambda_+] \\
&~~~ \Real{E_{\mathrm{noncoro},\lambda_0}(\vect{k})} \leq \Real{E_{\mathrm{noncoro}, \lambda_0}(\vect{0})} ~~ \forall~\vect{k} \\
&~~~ -\Real{E_{\mathrm{noncoro},\lambda_0}(\vect{k})} \leq \Real{E_{\mathrm{noncoro}, \lambda_0}(\vect{0})} ~~ \forall~\vect{k} \\
&~~~ \Imag{E_{\mathrm{noncoro},\lambda_0}(\vect{k})} \leq \Real{E_{\mathrm{noncoro}, \lambda_0}(\vect{0})} ~~ \forall~\vect{k} \\
&~~~ -\Imag{E_{\mathrm{noncoro},\lambda_0}(\vect{k})} \leq \Real{E_{\mathrm{noncoro}, \lambda_0}(\vect{0})} ~~ \forall~\vect{k}
\end{align}
\end{subequations}

Here $S_\mathrm{expected}$ is the expected transmission of the coronagraphic design. After optimization, this expected Strehl ratio can be updated by:
\begin{equation}
    S_\mathrm{expected} = (\Real{E_{\mathrm{noncoro}, \lambda_0}(\vect{0})})^2.
\end{equation}
The above optimization problem is then restarted with the updated expected Strehl ratio. This process is repeated until the expected Strehl ratio converges.

\bibliography{main}

\begin{thebibliography}{}
\expandafter\ifx\csname natexlab\endcsname\relax\def\natexlab#1{#1}\fi
\providecommand{\url}[1]{\href{#1}{#1}}

\bibitem[{{Beuzit} {et~al.}(2008){Beuzit}, {Feldt}, {Dohlen}, {Mouillet},
  {Puget}, {Wildi}, {Abe}, {Antichi}, {Baruffolo}, {Baudoz}, {Boccaletti},
  {Carbillet}, {Charton}, {Claudi}, {Downing}, {Fabron}, {Feautrier},
  {Fedrigo}, {Fusco}, {Gach}, {Gratton}, {Henning}, {Hubin}, {Joos}, {Kasper},
  {Langlois}, {Lenzen}, {Moutou}, {Pavlov}, {Petit}, {Pragt}, {Rabou}, {Rigal},
  {Roelfsema}, {Rousset}, {Saisse}, {Schmid}, {Stadler}, {Thalmann}, {Turatto},
  {Udry}, {Vakili}, \& {Waters}}]{beuzit2008sphere}
{Beuzit}, J.-L., {Feldt}, M., {Dohlen}, K., {et~al.} 2008, in \procspie, Vol.
  7014, Ground-based and Airborne Instrumentation for Astronomy II, 701418

\bibitem[{Borucki {et~al.}(2011)Borucki, Koch, Basri, Batalha, Brown, Bryson,
  Caldwell, Christensen-Dalsgaard, Cochran, DeVore,
  {et~al.}}]{borucki2011characteristics}
Borucki, W.~J., Koch, D.~G., Basri, G., {et~al.} 2011, The Astrophysical
  Journal, 736, 19

\bibitem[{Carlotti {et~al.}(2011)Carlotti, Vanderbei, \&
  Kasdin}]{carlotti2011optimal}
Carlotti, A., Vanderbei, R., \& Kasdin, N.~J. 2011, Opt. Express, 19, 26796.
\newblock \url{http://www.opticsexpress.org/abstract.cfm?URI=oe-19-27-26796}

\bibitem[{{Charbonneau} {et~al.}(2000){Charbonneau}, {Brown}, {Latham}, \&
  {Mayor}}]{Charbonneau2000}
{Charbonneau}, D., {Brown}, T.~M., {Latham}, D.~W., \& {Mayor}, M. 2000, \apjl,
  529, L45

\bibitem[{{Close} {et~al.}(2012){Close}, {Males}, {Kopon}, {Gasho}, {Follette},
  {Hinz}, {Morzinski}, {Uomoto}, {Hare}, {Riccardi}, {Esposito}, {Puglisi},
  {Pinna}, {Busoni}, {Arcidiacono}, {Xompero}, {Briguglio}, {Quiros- Pacheco},
  \& {Argomedo}}]{Close2012SPIEMagAO}
{Close}, L.~M., {Males}, J.~R., {Kopon}, D.~A., {et~al.} 2012, in Adaptive
  Optics Systems III, Vol. 8447, 84470X

\bibitem[{Codona {et~al.}(2006)Codona, Kenworthy, Hinz, Angel, \&
  Woolf}]{codona2006high}
Codona, J., Kenworthy, M., Hinz, P., Angel, J., \& Woolf, N. 2006, in SPIE
  Astronomical Telescopes+ Instrumentation, International Society for Optics
  and Photonics, 62691N--62691N

\bibitem[{Currie {et~al.}(2018)Currie, Kasdin, Groff, Lozi, Jovanovic, Guyon,
  Brandt, Martinache, Chilcote, Skaf, {et~al.}}]{currie2018laboratory}
Currie, T., Kasdin, N.~J., Groff, T.~D., {et~al.} 2018, Publications of the
  Astronomical Society of the Pacific, 130, 044505

\bibitem[{Doelman {et~al.}(2017)Doelman, Snik, Warriner, \&
  Escuti}]{doelman2017patterned}
Doelman, D.~S., Snik, F., Warriner, N.~Z., \& Escuti, M.~J. 2017, in Techniques
  and Instrumentation for Detection of Exoplanets VIII, Vol. 10400,
  International Society for Optics and Photonics, 104000U

\bibitem[{Esc{\'a}rate {et~al.}(2018)Esc{\'a}rate, Christou, Rahmer, Hill,
  Miller, \& Taylor}]{escarate2018vibration}
Esc{\'a}rate, P., Christou, J.~C., Rahmer, G., {et~al.} 2018, in Adaptive
  Optics Systems VI, Vol. 10703, International Society for Optics and
  Photonics, 107034F

\bibitem[{Fogarty {et~al.}(2018)Fogarty, Mazoyer, Laurent, Soummer, N'Diaye,
  Stark, \& Pueyo}]{fogarty2018optimal}
Fogarty, K., Mazoyer, J., Laurent, K.~S., {et~al.} 2018, in Space Telescopes
  and Instrumentation 2018: Optical, Infrared, and Millimeter Wave, Vol. 10698,
  International Society for Optics and Photonics, 106981J

\bibitem[{Fusco {et~al.}(2014)Fusco, Sauvage, Petit, Costille, Dohlen,
  Mouillet, Beuzit, Kasper, Suarez, Soenke, {et~al.}}]{fusco2014final}
Fusco, T., Sauvage, J.-F., Petit, C., {et~al.} 2014, in Adaptive Optics Systems
  IV, Vol. 9148, International Society for Optics and Photonics, 91481U

\bibitem[{{Gonsalves}(1982)}]{Gonsalves1982}
{Gonsalves}, R.~A. 1982, Optical Engineering, 21, 829

\bibitem[{Guerri {et~al.}(2011)Guerri, Daban, Robbe-Dubois, Douet, Abe,
  Baudrand, Carbillet, Boccaletti, Bendjoya, Gouvret,
  {et~al.}}]{guerri2011apodized}
Guerri, G., Daban, J.-B., Robbe-Dubois, S., {et~al.} 2011, Experimental
  Astronomy, 30, 59

\bibitem[{Gurobi~Optimization(2016)}]{gurobi}
Gurobi~Optimization, I. 2016, Gurobi Optimizer Reference Manual,  Gurobi
  Optimization, Inc.
\newblock \url{http://www.gurobi.com}

\bibitem[{{Henry} {et~al.}(2000){Henry}, {Marcy}, {Butler}, \&
  {Vogt}}]{Henry2000}
{Henry}, G.~W., {Marcy}, G.~W., {Butler}, R.~P., \& {Vogt}, S.~S. 2000, \apjl,
  529, L41

\bibitem[{Hunter(2007)}]{hunter2007matplotlib}
Hunter, J.~D. 2007, Computing In Science \& Engineering, 9, 90

\bibitem[{{Jovanovic} {et~al.}(2015){Jovanovic}, {Martinache}, {Guyon},
  {Clergeon}, {Singh}, {Kudo}, {Garrel}, {Newman}, {Doughty}, {Lozi}, {Males},
  {Minowa}, {Hayano}, {Takato}, {Morino}, {Kuhn}, {Serabyn}, {Norris},
  {Tuthill}, {Schworer}, {Stewart}, {Close}, {Huby}, {Perrin}, {Lacour},
  {Gauchet}, {Vievard}, {Murakami}, {Oshiyama}, {Baba}, {Matsuo}, {Nishikawa},
  {Tamura}, {Lai}, {Marchis}, {Duchene}, {Kotani}, \&
  {Woillez}}]{Jovanovic2015scexao}
{Jovanovic}, N., {Martinache}, F., {Guyon}, O., {et~al.} 2015, \pasp, 127, 890

\bibitem[{Kasdin {et~al.}(2003)Kasdin, Vanderbei, Spergel, \&
  Littman}]{kasdin2003extrasolar}
Kasdin, N.~J., Vanderbei, R.~J., Spergel, D.~N., \& Littman, M.~G. 2003, The
  Astrophysical Journal, 582, 1147.
\newblock \url{https://doi.org/10.1086%2F344751}

\bibitem[{Kushner(1964)}]{kushner1964new}
Kushner, H.~J. 1964, Journal of Basic Engineering, 86, 97

\bibitem[{{Macintosh} {et~al.}(2008){Macintosh}, {Graham}, {Palmer}, {Doyon},
  {Dunn}, {Gavel}, {Larkin}, {Oppenheimer}, {Saddlemyer}, {Sivaramakrishnan},
  {Wallace}, {Bauman}, {Erickson}, {Marois}, {Poyneer}, \&
  {Soummer}}]{Macintosh2008gpi}
{Macintosh}, B.~A., {Graham}, J.~R., {Palmer}, D.~W., {et~al.} 2008, in
  \procspie, Vol. 7015, Adaptive Optics Systems, 701518

\bibitem[{Males {et~al.}(2014)Males, Close, Morzinski, Wahhaj, Liu, Skemer,
  Kopon, Follette, Puglisi, Esposito, Riccardi, Pinna, Xompero, Briguglio,
  Biller, Nielsen, Hinz, Rodigas, Hayward, Chun, Ftaclas, Toomey, \&
  Wu}]{Males2014MagAO}
Males, J.~R., Close, L.~M., Morzinski, K.~M., {et~al.} 2014, The Astrophysical
  Journal, 786, 32.
\newblock \url{http://stacks.iop.org/0004-637X/786/i=1/a=32}

\bibitem[{{Mayor} \& {Queloz}(1995)}]{mayor1995jupiter}
{Mayor}, M., \& {Queloz}, D. 1995, \nat, 378, 355

\bibitem[{Mennesson {et~al.}(2016)Mennesson, Gaudi, Seager, Cahoy,
  Domagal-Goldman, Feinberg, Guyon, Kasdin, Marois, Mawet, Tamura, Mouillet,
  Prusti, Quirrenbach, Robinson, Rogers, Scowen, Somerville, Stapelfeldt,
  Stern, Still, Turnbull, Booth, Kiessling, Kuan, \&
  Warfield}]{mennesson2016habex}
Mennesson, B., Gaudi, S., Seager, S., {et~al.} 2016, in Space Telescopes and
  Instrumentation 2016: Optical, Infrared, and Millimeter Wave, Vol. 9904, 9904
  -- 9904 -- 10.
\newblock \url{https://doi.org/10.1117/12.2240457}

\bibitem[{Miller {et~al.}(2017)Miller, Guyon, \& Males}]{miller2017spatial}
Miller, K., Guyon, O., \& Males, J. 2017, Journal of Astronomical Telescopes,
  Instruments, and Systems, 3, 049002

\bibitem[{N'Diaye {et~al.}(2015)N'Diaye, Pueyo, \&
  Soummer}]{ndiaye2015apodized}
N'Diaye, M., Pueyo, L., \& Soummer, R. 2015, The Astrophysical Journal, 799,
  225

\bibitem[{{N'Diaye} {et~al.}(2016){N'Diaye}, {Soummer}, {Pueyo}, {Carlotti},
  {Stark}, \& {Perrin}}]{ndiaye2016apodized}
{N'Diaye}, M., {Soummer}, R., {Pueyo}, L., {et~al.} 2016, \apj, 818, 163

\bibitem[{Otten {et~al.}(2017)Otten, Snik, Kenworthy, Keller, Males, Morzinski,
  Close, Codona, Hinz, Hornburg, {et~al.}}]{otten2017sky}
Otten, G.~P., Snik, F., Kenworthy, M.~A., {et~al.} 2017, The Astrophysical
  Journal, 834, 175

\bibitem[{Pathak {et~al.}(2016)Pathak, Guyon, Jovanovic, Lozi, Martinache,
  Minowa, Kudo, Takami, Hayano, \& Narita}]{pathak2016high}
Pathak, P., Guyon, O., Jovanovic, N., {et~al.} 2016, Publications of the
  Astronomical Society of the Pacific, 128, 124404

\bibitem[{Por(2017)}]{por2017optimal}
Por, E.~H. 2017, in Techniques and Instrumentation for Detection of Exoplanets
  VIII, Vol. 10400, International Society for Optics and Photonics, 104000V

\bibitem[{Por {et~al.}(2018)Por, Haffert, Radhakrishnan, Doelman, Van~Kooten,
  \& Bos}]{por2018hcipy}
Por, E.~H., Haffert, S.~Y., Radhakrishnan, V.~M., {et~al.} 2018, in Proc.
  {{SPIE}}, Vol. 10703, Adaptive Optics Systems VI.
\newblock \url{https://doi.org/10.1117/12.2314407}

\bibitem[{Pueyo {et~al.}(2017)Pueyo, Zimmerman, Bolcar, Groff, Stark, Ruane,
  Jewell, Soummer, Laurent, Wang, Redding, Mazoyer, Fogarty, Juanola-Parramon,
  Domagal-Goldman, Roberge, Guyon, \& Mandell}]{pueyo2017luvoir}
Pueyo, L., Zimmerman, N., Bolcar, M., {et~al.} 2017, in UV/Optical/IR Space
  Telescopes and Instruments: Innovative Technologies and Concepts VIII, Vol.
  10398, 10398 -- 10398 -- 20.
\newblock \url{https://doi.org/10.1117/12.2274654}

\bibitem[{Riggs {et~al.}(2017)Riggs, Zimmerman, Nemati, \&
  Krist}]{riggs2017shaped}
Riggs, A.~E., Zimmerman, N.~T., Nemati, B., \& Krist, J. 2017, in Techniques
  and Instrumentation for Detection of Exoplanets VIII, Vol. 10400,
  International Society for Optics and Photonics, 104000O

\bibitem[{Ruane {et~al.}(2018)Ruane, Riggs, Mazoyer, Por, N’Diaye, Huby,
  Baudoz, Galicher, Douglas, Knight, {et~al.}}]{ruane2018review}
Ruane, G., Riggs, A., Mazoyer, J., {et~al.} 2018, in Space Telescopes and
  Instrumentation 2018: Optical, Infrared, and Millimeter Wave, Vol. 10698,
  International Society for Optics and Photonics, 106982S

\bibitem[{Slepian(1965)}]{slepian1965analytic}
Slepian, D. 1965, JOSA, 55, 1110

\bibitem[{Snik {et~al.}(2012)Snik, Otten, Kenworthy, Miskiewicz, Escuti,
  Packham, \& Codona}]{snik2012vector}
Snik, F., Otten, G., Kenworthy, M., {et~al.} 2012, in SPIE Astronomical
  Telescopes+ Instrumentation, International Society for Optics and Photonics,
  84500M--84500M

\bibitem[{Snoek {et~al.}(2012)Snoek, Larochelle, \& Adams}]{snoek2012practical}
Snoek, J., Larochelle, H., \& Adams, R.~P. 2012, in Advances in neural
  information processing systems, 2951--2959

\bibitem[{Soummer(2004)}]{soummer2004apodized}
Soummer, R. 2004, The Astrophysical Journal Letters, 618, L161

\bibitem[{{Spergel} {et~al.}(2013){Spergel}, {Gehrels}, {Breckinridge},
  {Donahue}, {Dressler}, {Gaudi}, {Greene}, {Guyon}, {Hirata}, {Kalirai},
  {Kasdin}, {Moos}, {Perlmutter}, {Postman}, {Rauscher}, {Rhodes}, {Wang},
  {Weinberg}, {Centrella}, {Traub}, {Baltay}, {Colbert}, {Bennett},
  {Kiessling}, {Macintosh}, {Merten}, {Mortonson}, {Penny}, {Rozo},
  {Savransky}, {Stapelfeldt}, {Zu}, {Baker}, {Cheng}, {Content}, {Dooley},
  {Foote}, {Goullioud}, {Grady}, {Jackson}, {Kruk}, {Levine}, {Melton},
  {Peddie}, {Ruffa}, \& {Shaklan}}]{spergel2013wfirst}
{Spergel}, D., {Gehrels}, N., {Breckinridge}, J., {et~al.} 2013, arXiv
  e-prints, arXiv:1305.5422

\bibitem[{Walt {et~al.}(2011)Walt, Colbert, \& Varoquaux}]{walt2011numpy}
Walt, S. v.~d., Colbert, S.~C., \& Varoquaux, G. 2011, Computing in Science \&
  Engineering, 13, 22

\bibitem[{Waterhouse(1983)}]{waterhouse1983symmetric}
Waterhouse, W.~C. 1983, The American Mathematical Monthly, 90, 378

\bibitem[{{Zimmerman} {et~al.}(2016){Zimmerman}, {Eldorado Riggs}, {Jeremy
  Kasdin}, {Carlotti}, \& {Vanderbei}}]{zimmerman2016shaped}
{Zimmerman}, N.~T., {Eldorado Riggs}, A.~J., {Jeremy Kasdin}, N., {Carlotti},
  A., \& {Vanderbei}, R.~J. 2016, Journal of Astronomical Telescopes,
  Instruments, and Systems, 2, 011012

\end{thebibliography}

\end{document}